\documentclass[12pt]{article}
\usepackage{amsmath}
\usepackage{graphicx}
\usepackage{enumerate}
\usepackage{natbib}
\usepackage{url} 

\newcommand{\blind}{1}

\usepackage[ruled,vlined,linesnumbered]{algorithm2e}
\usepackage[colorlinks = true, linkcolor = blue, urlcolor  = blue, citecolor = blue, anchorcolor = blue]{hyperref}
\usepackage{amsfonts}
\usepackage{amssymb}
\usepackage{booktabs}
\usepackage{bbm}
\usepackage{xr}
\RequirePackage{booktabs}
\RequirePackage{multirow}
\RequirePackage{enumerate}
\usepackage{mathrsfs}

\newtheorem{theorem}{Theorem}

\newtheorem{conjecture}{Conjecture}
\newtheorem{assumption}{Assumption}

\newtheorem{corollary}{Corollary}

\def\be{\begin{equation}}
\def\ee{\end{equation}}
\def\bea{\begin{eqnarray}}
\def\eea{\end{eqnarray}}
\def\bd{\begin{displaymath}}
\def\ed{\end{displaymath}}
\def\bda{\begin{eqnarray*}}
\def\eda{\end{eqnarray*}}
\def\bsm{\begin{small}}
\def\esm{\end{small}}
\def\T{{ \mathrm{\scriptscriptstyle T} }}

\newcommand{\beq}{\begin{eqnarray}}
\newcommand{\eeq}{\end{eqnarray}}

\newcommand{\V}{\rm Var}

\newcommand{\bI}{{\mathbf I}}

\newcommand{\bU}{{\mathbf U}}

\newcommand{\bX}{{\mathbf X}}
\newcommand{\bY}{{\mathbf Y}}
\newcommand{\bZ}{{\mathbf Z}}

\newcommand{\bt}{{\mathbf t}}

\newcommand{\bell} {\boldsymbol{\ell}}

\newcommand{\bSigma}{\boldsymbol{\Sigma}}

\newcommand{\bxi} {\boldsymbol{\xi}}
\newcommand{\bmu} {\boldsymbol{\mu}}

\newcommand{\bzero}{{\mathbf 0}}

\numberwithin{equation}{section}

\begin{document}

\def\spacingset#1{\renewcommand{\baselinestretch}%
{#1}\small\normalsize} \spacingset{1}


\if1\blind
{
  \title{\bf High-dimensional Clustering and Signal Recovery under Block Signals}
  \author{Wu Su \\
  Center for Big Data Research, Peking University, Beijing, China \\
  and \\
  Yumou Qiu \\
  School of Mathematical Sciences,
  Peking University, Beijing, China
  }
  \date{}
  \maketitle
} \fi

\if0\blind
{
  \bigskip
  \bigskip
  \bigskip
  \begin{center}
    {\LARGE\bf High-dimensional Clustering and Signal Recovery under Block Signals}
\end{center}
  \medskip
} \fi

\bigskip
\begin{abstract}

This paper studies computationally efficient methods and their minimax optimality for high-dimensional clustering and signal recovery under block signal structures. We propose two sets of methods, cross-block feature aggregation PCA (CFA-PCA) and moving average PCA (MA-PCA), designed for sparse and dense block signals, respectively. Both methods adaptively utilize block signal structures, applicable to non-Gaussian data with heterogeneous variances and non-diagonal covariance matrices. Specifically, the CFA method utilizes a block-wise U-statistic to aggregate and select block signals non-parametrically from data with unknown cluster labels. We show that the proposed methods are consistent for both clustering and signal recovery under mild conditions and weaker signal strengths than the existing methods without considering block structures of signals. 
Furthermore, we derive both statistical and computational minimax lower bounds (SMLB and CMLB) for high-dimensional clustering and signal recovery under block signals, where the CMLBs are restricted to algorithms with polynomial computation complexity. The minimax boundaries 
partition signals into regions of impossibility and possibility. No algorithm (or no polynomial time algorithm) can achieve consistent clustering or signal recovery if the signals fall into the statistical (or computational) region of impossibility.
We show that the proposed CFA-PCA and MA-PCA methods can achieve the CMLBs for the sparse and dense block signal regimes, respectively, indicating the proposed methods are computationally minimax optimal. 
A tuning parameter selection method is proposed based on post-clustering signal recovery results. Simulation studies are conducted to evaluate the proposed methods. 
A case study on global temperature change demonstrates their utility in practice. 
\end{abstract}

\noindent%
\vfill

\newpage
\spacingset{1.9} 

\section{Introduction}

Grouping observations into clusters is an important problem in data analysis, and clustering is a key method of unsupervised learning. In many scientific studies, the signals that separate different clusters are likely to appear in contiguous locations, forming blocks of signals.
For example, \cite{Chen2024} employed hierarchical clustering on high-resolution satellite data to examine global precipitation patterns, uncovering noteworthy diurnal cycles and seasonal variations across diverse areas, where spatially contiguous regions capture distinct climate patterns. 
Other examples of data with block signals include medical imaging data, neuroimaging data, copy number variation (CNV) data in genomics, global geoscience data \citep{Jeng2010, Feng2012}.
Utilizing the block structures of signals can lead to more accurate clustering results. In this paper, we develop minimax optimal procedures for clustering and signal recovery for high-dimensional data with block signals under the polynomial-time computational constraint. 

There are several recent advances on 
high-dimensional clustering. 
\cite{Liu2008} considered detecting clusters in high-dimensional data. \cite{Sun2012} considered regularized $k$-means clustering. \cite{Cai2019} proposed an expectation-maximization (EM) algorithm for high-dimensional Gaussian mixtures, demonstrating its optimality in convergence rates. 
Compared to model-based EM algorithms that rely on parametric distribution assumption and $k$-means method that has a high computational complexity, spectral clustering has gained more attention in high-dimensional data analysis recently. 
\cite{Loffler2021} established the optimal convergence rate of spectral clustering in Gaussian mixture models with isotropic covariance matrix.
\cite{Abbe2022, Zhang2024} developed the leave-one-out analysis framework for eigen-analysis and spectral clustering. Also see spectral clustering for social network data in \cite{jin2015, Huang2023}.
Although the aforementioned methods were shown effective and optimal
in some settings, they didn't consider the phase transition problem of clustering which requires minimax analysis for a finer signal specification. 
Especially, the problems of clustering and signal recovery for data with block (or spatially contiguous) signals 
have not been well studied. 

This paper adopts a minimax framework to investigate the fundamental limits of clustering in terms of signal strength, sparsity, and block size. \cite{Ingster1997, Donoho2004} first derived the detection boundary for testing means
under a Bayesian setting of randomly located signals, which partitions the parameter space of signal strength and sparsity into the regions of detectable and undetectable. They showed that no test can separate the null and alternative hypotheses in the undetectable region. \cite{Kou2022} extended this result to test block-structured means. \cite{Chen2023, qiu-2024} derived the minimax detection boundaries for testing high-dimensional covariance and precision matrices under non-block signals. For clustering, \cite{Jin2016} introduced the influential feature principal component analysis (IF-PCA) under sparse non-block signals, and subsequently, \cite{Jin2017} established the statistical clustering boundary under non-block signals. 
However, this statistical boundary may not be attainable by a polynomial-time clustering algorithm
in some regions of signals. 
It is important to understand the limits of clustering algorithms that can be computed in polynomial time.

Some high-dimensional inference tasks are inherently difficult from a computational point of view. The low-degree polynomial conjecture \citep{Hopkins2018, Kunisky2022} has recently gained attention for its ability to distinguish what is statistically possible from what can be computed in polynomial time. It helps to understand the gap between statistical performance and computational feasibility. Intuitively, if no polynomial functions can detect the clustering signals, then successful clustering is not achievable by any polynomial-time algorithm. Although not rigorously proven, substantial evidence supporting it was collected in \cite{Hopkins2018}. \cite{Loffler2022} used this conjecture to establish the polynomial-time minimax optimal convergence rate for sparse clustering. 
But, they didn't study the minimax results for signal recovery and neither considered block signals. \cite{Lyu2023} applied the conjecture to reveal a statistical-computational gap for estimating the low-rank mean of a two-component Gaussian mixture model.

In this paper, we propose two sets of methods for clustering and signal recovery under sparse and dense block signals, respectively. For sparse signals, inspired by IF-PCA \citep{Jin2016}, we first perform feature selection and then cluster the selected features. To handle non-Gaussian distributions and heterogeneous variances, we propose a new method, called the cross-block feature aggregation (CFA) procedure, to aggregate and filter out useful features from clustered data based on pairwise U-statistics. The proposed CFA method prevents signal cancellation from unknown cluster labels and avoids estimating unknown variances. It works under non-Gaussian distributions and is more general than IF-PCA designed for the Gaussian distribution with diagonal covariance. After identifying the informative blocks, we perform spectral clustering on the aggregated features of the selected blocks, resulting in the CFA-PCA algorithm. 
For dense signals, we propose the moving average PCA (MA-PCA) procedure. Difference from CFA-PCA, MA-PCA first applies spectral clustering to aggregated features by moving average. Based on the clustering results, we propose a post-clustering signal recovery algorithm for feature selection, inspired by spatial scan statistics \citep{Arias-Castro2005, Jeng2010}. A tuning parameter selection method is proposed based on post-clustering signal recovery results. 

We prove that our methods are consistent for both clustering and signal recovery under mild conditions, weaker than the existing methods for sparse clustering and block signal identification. 
Our results highlight the advantages of the proposed methods in exploiting block structures of signals for both sparse and dense regimes, which are consistent under weaker signal strengths than the existing methods without considering block signals. Simulation studies show that our methods provide more accurate estimates than existing methods. A case study on global temperature data demonstrates their utility. 

Furthermore, we establish both the statistical and computational minimax lower boundaries (SMLB and CMLB) for clustering and signal recovery in a high-dimensional setting with block signals. 
The SMLB and CMLB partition the parameter space of signal sparsity, strength and block size into the statistical and computational regions of impossibility and possibility, respectively.
No algorithm (or no polynomial time algorithm) can achieve consistent clustering or signal recovery if the signals fall into the statistical (or computational) region of impossibility. Compared to \cite{Loffler2022}, our results derive the phase transition diagram under block signals, show the gap between SMLB and CMLB, and provide the minimax results for signal recovery. We show that the proposed CFA-PCA and MA-PCA methods can achieve the CMLBs for the sparse and dense block signal regimes, respectively, demonstrating the proposed methods are computationally minimax optimal.

The remainder of this paper is organized as follows. Section \ref{sec:settings} presents the problem setting and the block signal structure. 
Sections \ref{sec:upperbound_sparse} and \ref{sec:upperbound_dense} introduce the proposed CFA-PCA and MA-PCA procedures. 
Section \ref{sec:lowerbound} derives the statistical and computational minimax lower boundaries for clustering and signal recovery, and reveals the computational minimax optimality of the proposed methods.  
Section \ref{sec:ext} extends the proposed methods to tensor-valued data. Section \ref{sec:exp} and \ref{sec:data} report the simulation and case studies, respectively. Technical proofs are relegated to the Supplementary Material (SM).

\section{Problem setting} \label{sec:settings}

We consider the two‐group high-dimensional clustering setting with block signals. 
To simplify notations and concentrate on main contributions, we first present our results under vector-valued data, where the variables have a natural ordering in a sequence. Extensions to tensor-valued data are discussed in Section \ref{sec:ext}.
Let $\{\bX_1, \ldots, \bX_n\}$ denote a sample of $p$-dimensional independent observations of size $n$, where $\bX_i = (X_{i1}, \ldots, X_{ip})^{\T}$. 
Here, the dimension can be much larger than the sample size. Following the high-dimensional clustering model proposed in \cite{Jin2017}, we assume the cluster label of each observation is independent and identically distributed (IID) from a Bernoulli distribution. Let $\ell_i \sim 2\mbox{Bernoulli}( \varpi / (1 + \varpi) ) - 1$, where $\varpi$ is a positive constant indicating the proportion of the two groups.
Let $\bmu_1 = \mathbb{E}(\bX_i | \ell_i = 1)$ and $\bmu_{-1} = \mathbb{E}(\bX_i | \ell_i = -1)$ be the means of the two groups. Note that $\mathbb{E}(\bX_i) = (\varpi \bmu_1 + \bmu_{-1}) / (1 + \varpi)$ is the overall mean of the sample. We assume $\varpi \bmu_1 + \bmu_{-1} = \bzero$ for simplicity of presentation.

By a reparameterization of the mean and the cluster labels, we consider the model 
\be  \label{eq:mixgauss}
\bX_i = \tilde{\ell}_i \bmu + \bZ_i, \quad \bZ_i \stackrel{\text{IID}}{\sim} F \text{ with mean zero and covariance } \bSigma,
\ee 
and $\tilde{\ell}_i$ is IID from the binary distribution
\be
\tilde{\ell}_i = 
\left\{
\begin{array}{cc}
 1,      & \mbox{with probability $\varpi / (1 + \varpi)$}, \\
-\varpi, & \mbox{with probability $1 / (1 + \varpi)$},
\end{array}
\right.
\ee 
for $1 \le i \le n$, where $\bmu = (\mu_1, \ldots, \mu_p)^\T \in \mathbb{R}^p$ is an unknown mean vector and $\bSigma$ is a $p \times p$ unknown covariance matrix.
Here, $\tilde{\ell}_i$ can be viewed as a generalized cluster label, where $\operatorname{sign}(\tilde{\ell}_i) = \ell_i$ is the label of the $i$th observation and $\operatorname{sign}(\cdot)$ denotes the sign function. 
Under this setting, we investigate both clustering and signal recovery problems, which aim to identify the class labels $\bell = (\ell_1, \ldots, \ell_n)^\T$ and to recover the support of the signal, defined as $S(\bmu) = \{ j : \mu_j \neq 0 \}$, respectively.

%

We consider a block structure for the mean \(\bmu\). Its nonzero components are distributed across non-overlapping ordered blocks \(\mathbb{I}_0 = \{I_{0,g}: 1\leq g \leq m\}\), where $I_{0,g} = \{j_g, j_g + 1, \ldots, j_g + l_g\}$ and $j_g + l_g < j_{g + 1}$ for all $g$, $l_g$ denotes the length of $g$th block, and \(m\) denotes the number of blocks. Let $d_0 = \min_g\{j_{g + 1} - (j_g + l_g)\}$ be the minimum distance between blocks, $d_g = \max\{j_g - (j_1 + l_1), j_m - (j_g + l_g)\}$ be the maximum distance from $I_{0,g}$ to the rest $(m-1)$ blocks, and $\tilde{d} = \min_{1 \leq g \leq m} d_g$. It is trivial to notice that $\tilde{d} \geq m d_0 / 2$. 
The number of nonzero entries of $\bmu$ is denoted as $s = |S(\bmu)| = \sum_{g=1}^m |I_{0,g}|$, where $|A|$ denotes the cardinality of a set $A$. 

In this paper, 
we consider the setting of expanding regions that $m \to \infty$ and the block length $|I_{0,g}| = \rho_{g} b$, where $\{\rho_g\}$ are positive constants and $b$ is positive and can depend on $p$. Our setting covers the case of diverging block length where $b$ diverges to infinity as $p \to \infty$, for example, $b = p^{\alpha}$ for $\alpha \in (0, 1)$. It also covers the fixed block length and non-block signal settings, which correspond to $b = 1$. 
Note that the total number of non-overlapping blocks of size $b$ is at the order $p / b$. Regarding the number of blocks with signals, we call the settings as sparse and dense block signal regimes when $m \ll \sqrt{p / b}$ and $m \gg \sqrt{p / b}$, respectively. 
Such definitions of sparse and dense regimes for block signals are similar to those for the non-block signal setting, which uses $\sqrt{p}$ to separate the two regimes \citep{Donoho2004, Jin2017}. We will develop computationally minimax optimal procedures for both clustering and signal recovery under sparse and dense block signals, respectively.

We impose the following regularity conditions on the block signal structure and data.

\begin{assumption}\label{assu:sub_gaussian} 
For any \(\bt \in \mathbb{R}^p\) and a constant $C > 0$, $\mathbb{E}\{ \exp(\bt^{\T} \bZ_i) \} \leq \exp(C \bt^{\T}\bSigma \bt/ 2)$. 
\end{assumption} 

\begin{assumption}\label{assu:separable} 
The minimum distance between blocks satisfies ${d_0} / (\max_{1\le g\le m} |I_{0,g}|) \rightarrow \infty$. 
\end{assumption} 

\begin{assumption}\label{assu:strength} 
There exists $\mu_0 > 0$ and $c_0 \geq 1$ such that $c_0^{-1}\mu_0 \le |\mu_j| \le c_0 \mu_0$ for any $j \in S(\bmu)$, and $\operatorname{sign}(\mu_j)=e_g$ for $j \in I_{0,g}$ and $e_g = \pm 1$. 
\end{assumption} 



Assumption \ref{assu:sub_gaussian} assumes the data are sub-Gaussian distributed, which is commonly made in high-dimensional literature \citep{Bickel2008, Cai2010, Cai2023}.
Assumption \ref{assu:separable} assumes that all blocks of signals are asymptotically separable, meaning the distance between any pair of signal blocks is much larger than their sizes. This assumption is needed for the correct recovery of each block. Intuitively, if the distance between two signal blocks is much smaller than their sizes, they can be viewed as one block asymptotically. A similar condition was made in \cite{Jeng2010} for signal identification under IID data without unknown clusters. If the left ends of all blocks, $\{j_1, \ldots, j_m\}$, are randomly selected from $\{1, \ldots, p\}$, we show in the SM that 
\be\label{eq:distance}
d_0 \geq 2 p m^{-2} \log^{-1}(p) \mbox{ \ and \ }
\tilde{d} \geq p m^{-1} \log^{-1}(p)
\ee
with probability converging to 1. It follows that $d_0 \gg b$ if $m = o((p/b)^{1/2} \log^{-1/2}(p))$, and Assumption \ref{assu:separable} is satisfied in the sparse block signal regime.
Assumption \ref{assu:strength} assumes the nonzero means are at the same order, but they can vary over $j \in S(\bmu)$. Within each block, the nonzero means have the same sign. Such block signal structures are commonly observed in geoscience and medical imaging data.
\cite{Kou2022} considered a similar block signal setting for testing high-dimensional means. However, they studied the setting of equal signal for all dimensions, which corresponds to $c_0 = 1$ in Assumption \ref{assu:strength}.    
We focus on two-sided signals rather than one-sided signals, and the signal strengths can be different on different components.

Our setting and assumptions for high-dimensional clustering and signal recovery under block signals are relatively general. 
Most of the existing high-dimensional clustering methods with sparsity constraints require the Gaussianity assumption and don't consider block signals \citep{Jin2016, Cai2019, Loffler2022}. 
\cite{Jeng2010, Kou2022} also assumed Gaussian data with identity covariance for block signal detection and identification. In contrast, our setting relaxes those assumptions, and the proposed algorithms in Sections \ref{sec:upperbound_sparse} and \ref{sec:upperbound_dense}  can be easily extended to tensor data, making them applicable to geoscience applications.

In the following, we introduce some notations used in the paper. For positive sequences $\{a_{1,n}\}$ and $\{a_{2,n}\}$, $a_{1,n} \asymp a_{2,n}$ denotes the two sequences being at the same order such that $c_1 \leq a_{1,n} / a_{2,n} \leq c_2$ for all $n$ and two positive constants $c_1$ and $c_2$. We use $a_{1,n} \lesssim a_{2,n}$ to denote $a_{1,n} / a_{2,n} \leq c_2$ for all $n$. 
Let $\bX = (\bX_1, \ldots, \bX_n)^{\T} = (\bX_{(1)}, \ldots, \bX_{(p)})$ denote the data matrix, where $\bX_{(j)} = (X_{1j}, \ldots, X_{nj})^{\T}$ is the vector of the $j$th covariate. 
Let $\mathbbm{1}(\cdot)$ denote the indicator function,
and $C$ denote a positive constant which may change from case to case. Let $\|\cdot\|$ and $\|\cdot\|_{\infty}$ denote the vector Euclidean and maximum norms, and $\|\cdot\|_2$ and $\|\cdot\|_{\mathrm{F}}$ denote the matrix $L_2$ and Frobenius norms, respectively.

\section{Methods for sparse block signals} \label{sec:upperbound_sparse} 

In this section, we introduce the polynomial-time algorithms for clustering and signal recovery under sparse block signals, where $m \ll \sqrt{p / b}$.
The minimax results in Section \ref{sec:lowerbound} show that consistent clustering requires stronger signals than signal recovery for sparse block signals, and vice versa for dense block signals. This indicates that we should conduct pre-clustering signal recovery first and cluster the observations based on the selected features for sparse block signals. Instead, for dense block signals, clustering should be applied first followed by a post-clustering signal recovery procedure. Therefore, the minimax optimal algorithms are different for sparse and dense signals. 


To handle clustered data with sparse block signals, we need to propose a new method for selecting useful features.
Let $\mathbb{J}_p(h_1) = \{(j, \ldots, j + h): 1 \leq h \leq h_1, 1 \leq j < j + h \leq p\}$ be the collection of all blocks with length no larger than $h_1$, where $|\mathbb{J}_p(h_1)| \leq p h_1$.
For $I \in \mathbb{J}_p(h_1)$, let $X_i(I) = \sum_{j \in I} X_{ij} / \sqrt{|I|}$, $Z_i(I) = \sum_{j \in I} Z_{ij} / \sqrt{|I|}$, $\mu(I) = \sum_{j \in I} \mu_j / \sqrt{|I|}$ and $\sigma^2(I) = {\V}\{Z_i(I)\}$, where $X_i(I) = \tilde{\ell}_i \mu(I) + Z_i(I)$. 
Sparse feature selection for clustered data is difficult as the signals in each observation are canceled in the overall mean $n^{-1}\sum_{i = 1}^{n} X_i(I)$ of all observations. 
Note that the $\chi^2$ statistic $n^{-1/2} \sum_{i=1}^n X_i^2(I)$ cannot identify the nonzero $\mu(I)$ as well if $\sigma^2(I)$ is unknown. 
To solve this problem,
\cite{Jin2016} used Kolmogorov-Smirnov (KS) statistic for feature screening under Gaussian mixture distributions with diagonal covariance and non-block signals, which compares column-wise empirical cumulative distribution functions (CDF) with the standard normal CDF. However, this method requires the Gaussian distribution assumption.  
This paper addresses feature selection problem under non-Gaussian distributions, block signals, and bandable covariance matrices, defined in \eqref{eq:bandable}. 

To remove the variance term in the $\chi^2$ statistic, we introduce a pairwise screening procedure.  
For $I_1, I_2 \in \mathbb{J}_p(h_1)$, consider the cross-product $ W_i(I_1, I_2) = X_i(I_1) X_i(I_2)$. Let $\sigma(I_1, I_2) = \mathbb{E}\{ Z_i(I_1) Z_i(I_2) \}$.
The advantages of using cross-products for signal screening can be reflected through their expectations ${\mathbb{E}}\{W_i(I_1, I_2)\} = \tilde{\ell}_i^2 \mu(I_1) \mu(I_2) + \sigma(I_1, I_2)$, which have the same sign over all observations so that signals would not be canceled by taking the average of $\{W_i(I_1, I_2)\}$. 
Furthermore, those expectations do not involve the variances $\sigma^2(I_1)$ and $\sigma^2(I_2)$. 
If $\sigma(I_1, I_2)$ is small, we can use the average $\overline{W}(I_1, I_2) = n^{-1/2} \sum_{i = 1}^{n} W_i(I_1, I_2)$ to test for $\mu(I_1) \mu(I_2) = 0$, and hence, to identify the blocks with nonzero means.

For data with a natural ordering or collected in spatial locations such as those in earth science problems, the dependence between two variables decreases as their distance increases. The following ``bandable'' class \citep{Bickel2008, Cai2010} describes such covariance matrices for variables observed in a sequence:
\be \label{eq:bandable}
\mathfrak{U}(\gamma, C) = \bigg\{ \bSigma = (\sigma_{j_1j_2}): |\sigma_{j_1j_2}|  \le C k^{-(\gamma + 1)} \text{ for all } k > 0 \bigg\}.  
\ee
This bandable class can be extended to spatial data that encompasses a wide range of spatial covariance functions \citep{
Sun2024}. Under the class in \eqref{eq:bandable}, $\sigma(I_1, I_2) \to 0$ as the distance between $I_1$ and $I_2$ diverges. Note that $\gamma$ can be treated as $\infty$ if $\bSigma$ is diagonal. 
Therefore, we consider the pair of blocks that are far apart. 

Let $\mathcal{E}_p(\{ j, \ldots, j + \tilde{c}_1 \}, \tilde{c}_2)$ be the expanding operator that expands the interval $\{ j, \ldots, j + \tilde{c}_1 \}$ from both ends by a window $\tilde{c}_2$, where
\be \label{eq:expand_operator}
\mathcal{E}_p( \{ j, \ldots, j + \tilde{c}_1 \}, \tilde{c}_2 ) = \{ j - \tilde{c}_2, \ldots, j + \tilde{c}_1 + \tilde{c}_2\} \cap \{ 1, \ldots, p \},
\ee
and $\tilde{c}_1, \tilde{c}_2 > 0$. For $I_1 \in \mathbb{J}_p(h_1)$ and a window size $h_2 > 0$, we consider its cross products with blocks that are at least $h_2$ apart. Let
\begin{equation} \label{eq:pair_segment}
\tilde{I}_1 = \underset{ I_2 \in \mathbb{J}_p(h_1), \ I_2 \cap \mathcal{E}(I_1, h_2) = \emptyset}{\operatorname{argmax}} \ |\overline{W}(I_1, I_2)| \mbox{ \ and \ }
W_0(I_1) = \overline{W}(I_1, \tilde{I}_1)
\end{equation} 
be the largest cross-product with $I_1$ and the corresponding block, satisfying $\tilde{I}_1 \cap \mathcal{E}(I_1, h_2) = \emptyset$.
We use $W_0(I)$ to detect signals in the block $I$. Let $\hat{\sigma}_{\mathrm{w}}^2(I) = n^{-1} \{ \sum_{i=1}^n W_i^2(I, \tilde{I}) - W_0^2(I) \}$ 
be the sample variance of $\{W_i(I, \tilde{I})\}_{i = 1}^{n}$.
We identify the significance blocks if their standardized aggregated average $|W_0(I)| / \hat{\sigma}_{\mathrm{w}}(I)$ is larger than the threshold $t_{\mathrm{w}, p} = \sqrt{6 \log(ph_1)}$. Let $\mathbb{Q}_{\mathrm{w}}^{(1)} = \big\{ I \in \mathbb{J}_p(h_1) : {|W_0(I)|} / {\hat{\sigma}_{\mathrm{w}}(I)} > t_{\mathrm{w}, p} \big\}$ 
be the set of all identified blocks with nonzero mean $\mu(I)$.
We select the first identified block as the largest $|{W}_0(I)|$ from $\mathbb{Q}^{(1)}$, denoted as $\hat{I}_1 = \operatorname{argmax}_{I \in \mathbb{Q}^{(1)}} |W_0(I)|$, and search the next significant block in the set $\mathbb{Q}^{(2)} = \mathbb{Q}^{(1)} \setminus \big\{ I \in \mathbb{Q}^{(1)} : I \cap \mathcal{E}_p(\hat{I}_1, \lfloor h_1 / 2 \rfloor) \neq \emptyset \big\}$
by removing any block intersecting an expanded block of $\hat{I}_1$ with length $|\hat{I}_1| + h_1$ from $\mathbb{Q}^{(1)}$. 
We continue this procedure by finding $\hat{I}_2 = \operatorname{argmax}_{I \in \mathbb{Q}^{(2)}} |W_0(I)|$ and constructing $\mathbb{Q}^{(3)} = \mathbb{Q}^{(2)} \setminus \big\{ I \in \mathbb{Q}^{(2)} : I \cap \mathcal{E}_p(\hat{I}_2, \lfloor h_1 / 2 \rfloor) \neq \emptyset \big\}$, until $\mathbb{Q}^{(k)}$ becomes empty. The condition $I \cap \mathcal{E}_p(\hat{I}_k, \lfloor h_1 / 2 \rfloor) = \emptyset$ in each step ensures that all identified blocks are at least $h_1 / 2$ separated from each other, guaranteeing the consistency of the number of identified blocks. We call this procedure as the cross-block
feature aggregation (CFA) method for pre-clustering signal identification.

\begin{algorithm}\label{alg:2}
	\caption{Cross-block feature aggregation PCA (CFA-PCA)}
	\label{alg:Pre-hoc-sgement}
	\KwIn{observation $\bX_1, \bX_2, \ldots, \bX_n$; window sizes $h_1, h_2$. }
	\KwOut{block identification $\hat{\mathbb{I}}_{\mathrm{cfa}} = \{\hat{I}_1, \ldots, \hat{I}_{\hat{m}}\}$ and estimated cluster labels $\hat{\bell}_{\mathrm{cfa}}$. }  
	\BlankLine
        
        Obtain $W_0(I)$ and $\hat{\sigma}_{\mathrm{w}}(I)$ for all $I \in \mathbb{J}_p(h_1)$ by \eqref{eq:pair_segment};
        
        $k=1$ and $\mathbb{Q}_{\mathrm{w}}^{(1)} = \big\{ I \in \mathbb{J}_p(h_1) : {|W_0(I)|} / {\hat{\sigma}_{\mathrm{w}}(I)} > t_{\mathrm{w}, p} \}$;

	\While{${\mathbb{Q}}^{(k)} \neq \emptyset$}{
        $\hat{I}_k =   {\operatorname{argmax}}_{I\in {\mathbb{Q}}^{(k)}}\, |W_0(I)|$;  $\mathbb{Q}^{(k+1)} = \mathbb{Q}^{(k)} \backslash \{ I\in \mathbb{Q}^{(k)}: I \cap \mathcal{E}_p(\hat{I}_k, \lfloor h_1/2 \rfloor)\ne \emptyset\}$; 
	}
    
Obtain $\hat{\mathbb{I}}_{\mathrm{cfa}} = \{\hat{I}_1, \ldots, \hat{I}_{\hat{m}}\}$ and $\hat{\bY}^{(\mathrm{cfa})}_{(k)} = |\hat{I}_k|^{-1/2} {\sum}_{j \in \hat{I}_k} \bX_{(j)}$ 
for $k = 1, \ldots, \hat{m}$;

Apply PCA on $\hat{\bY}^{(\mathrm{cfa})}$ to obtain the estimated cluster labels $\hat{\bell}_{\mathrm{cfa}}$ in \eqref{eq:ifa-pca}.
\end{algorithm}

Let $\hat{\mathbb{I}}_{\mathrm{cfa}} = \{\hat{I}_1, \hat{I}_2, \ldots, \hat{I}_{\hat{m}}\}$ denote the $\hat{m}$ identified blocks by the proposed CFA method. Then, we perform clustering solely on the aggregated mean of each selected block. Let $\hat{\bY}^{(\mathrm{cfa})} = (\hat{\bY}^{(\mathrm{cfa})}_{(1)}, \ldots, \hat{\bY}^{(\mathrm{cfa})}_{(\hat{m})})$ be the aggregated data matrix of the selected features, where $\hat{\bY}^{(\mathrm{cfa})}_{(k)} = |\hat{I}_k|^{-1/2} {\sum}_{j \in \hat{I}_k} \bX_{(j)}$ 
for $k = 1, \ldots, \hat{m}$. We apply PCA on $\hat{\bY}^{(\mathrm{cfa})}$ to cluster observations in the sparse signal regime. 
The cluster labels $\bell$ can be estimated by
\begin{equation} \label{eq:ifa-pca}
\hat{\bell}_{\mathrm{cfa}} = \operatorname{sign}(\hat{\bxi}_{\mathrm{cfa}}),
\end{equation}
where $\hat{\bxi}_{\mathrm{cfa}}$ is the first eigenvector of the matrix $\hat{\bY}^{(\mathrm{cfa})} \hat{\bY}^{(\mathrm{cfa}) \T}$. We call this clustering method cross-block feature aggregation PCA (CFA-PCA), outlined in Algorithm \ref{alg:2}.
%
The following theorem shows the consistency of $\hat{\bell}_{\mathrm{cfa}}$ under a suitable condition of the signal strength.


\begin{theorem} \label{th:UB_comp_sparse_cluster} 
For the clustering problem under the model in \eqref{eq:mixgauss}, under Assumptions \ref{assu:sub_gaussian}-\ref{assu:strength}, $\bSigma \in \mathfrak{U}(\gamma, C)$ in \eqref{eq:bandable}, $\max \{ \| \bSigma \|_2, \| \bSigma^{-1} \|_2\} \le C$ and $\sqrt{n} b \tilde{d}^{-\gamma-1} = o(1)$, if $h_1 = c_1 b$ and $h_2 = c_2 \tilde{d}$ for some constants $c_1 \geq \max_{g} \rho_g$ and $c_2 \in (0, 1)$, and $\bmu$ satisfies the signal-to-noise ratio condition 
\be \label{eq:signal_to_noise_pre}
\sqrt{n}{m}^{-1} \|\bmu\|^2 > c_3 \log p, \quad 
\min\{ (n / m)^{1/2}, 1 \} \| \bmu\|^2  > c_3 \log n
\ee
for a positive constant $c_3$, 
then we have 
$\min\{\|\hat{\bell}_{\mathrm{cfa}} + \bell\|_{\infty}, \|\hat{\bell}_{\mathrm{cfa}} - \bell\|_{\infty}\} \to 0$ with probability approaching to $1$ as $n, p \to \infty$. 
\end{theorem} 

Theorem \ref{th:UB_comp_sparse_cluster} establishes the consistency of the proposed CFA-PCA procedure for clustering under the bandable covariances with bounded eigenvalues. 
The first signal-to-noise condition in \eqref{eq:signal_to_noise_pre} is for block feature selection, which ensures the expectation $\sqrt{n}\mu(I_1) \mu(I_2)$ of the cross-products $\sqrt{n}W_i(I_1, I_2)$ diverges to infinity if $I_1$ and $I_2$ are the true signal blocks. The second signal-to-noise condition on $\min\{ (n / m)^{1/2}, 1 \} \| \bmu\|^2$ in \eqref{eq:signal_to_noise_pre} is much weaker than the classical condition $\min\{ (n / p)^{1/2}, 1 \} \| \bmu \|^2 \to \infty$ for spectral clustering without feature selection \citep{Cai2018, Abbe2022}. We have shown in the SM that $|\hat{I}_g| \asymp |I_{0,g}|$ for the sizes of the selected blocks by the proposed CFA procedure. This reduces the effective dimension from $p$ to $m$ in the CFA-PCA procedure for clustering. 
The condition $\sqrt{n} b \tilde{d}^{-\gamma-1} = o(1)$ ensures 
$\sqrt{n} \sigma(I_1, I_2) = o(1)$ for any pair of blocks that are at least $\tilde{d}$ apart, guaranteeing that no blocks will be falsely detected by the CFA procedure. Under \eqref{eq:distance} and $b = p^{\alpha}$, a sufficient condition for $\sqrt{n} b \tilde{d}^{-\gamma-1} = o(1)$ in the sparse block signal regime is $\gamma > \alpha / (\alpha + 1)$, which places no restriction on $\gamma$ when $\alpha = 0$. 
We provide a data-driven tuning parameter selection method for $h_1$ and $h_2$ in the simulation section.

\cite{Loffler2022} proposed a sparse spectral clustering method without considering block signals for Gaussian data with identity covariance. Their method is related to sparse PCA \citep{Zou2018}. It is consistent under the same signal-to-noise ratio condition as \eqref{eq:signal_to_noise_pre} with $b = 1$ for non-block signals. As a comparison, the proposed CFA-PCA method achieves consistent clustering for sparse non-block signals under weaker conditions on the data distribution and the covariance structure. 

The following theorem shows that the selected blocks $\hat{\mathbb{I}}_{\mathrm{cfa}}$ by the proposed CFA method is consistent to the true set of signals $\mathbb{I}_0$, if the signal strengths within each block are the same. Let $D(I_1, I_2)=1 - |I_1 \cap I_2| (|I_1| |I_2|)^{-1/2}$ be the dissimilarity between sets $I_1$ and $I_2$. 

\begin{theorem} \label{th:UB_comp_sparse_signal} 
For the signal recovery problem under the model in \eqref{eq:mixgauss}, under Assumptions \ref{assu:sub_gaussian}-\ref{assu:strength}, $\bSigma \in \mathfrak{U}(\gamma, C)$ in \eqref{eq:bandable}, $\max \{ \| \bSigma \|_2, \| \bSigma^{-1} \|_2\} \le C$, $\sqrt{n} b\tilde{d}^{-\gamma-1} = o(1)$, $\sqrt{n} b \mu_0^2 > c_3 \log p$ for $c_3 > 0$ and additionally $\mu_j = \tau_g$ for $j \in I_{0,g}$ and $g = 1, \ldots, m$, if $h_1 = c_1 b$ and $h_2 = c_2 \tilde{d}$ for some constants $c_1 \geq \max_{g} \rho_g$ and $c_2 \in (0, 1)$, 
then we have that $\mathbb{P}(\hat{m} = m) \to 1$ and $\mathbb{P}( \operatorname{max}_{I \in \mathbb{I}_0} \operatorname{min}_{\hat{I} \in \hat{\mathbb{I}}_{\mathrm{cfa}}} D(\hat{I}, I) > \zeta ) \to 0$ for any $\zeta > 0$. 
\end{theorem} 

Compared to Theorem \ref{th:UB_comp_sparse_cluster}, consistent signal recovery requires an additional condition that the means of all variables in each signal block $I_{0, g}$ are the same. \cite{Arias-Castro2005, Jeng2010} also require this condition for block signal identification. Without this condition, we may only identify a subset $\hat{I}_g$ of $I_{0, g}$ satisfying $|\hat{I}_g| \asymp |I_{0,g}|$, which is sufficient for consistent clustering.

\section{Methods for dense block signals} \label{sec:upperbound_dense} 

In this section, we consider clustering and signal recovery for dense block signals where $m \gg \sqrt{p/b}$. Different from the sparse signal regime, we first cluster the observations and then identify the components with signals based on the estimated clusters in the dense signal regime. 

To utilize the block signal structure, we conduct clustering on the aggregated variables by
moving average with window size $h_3$. Let 
$$
\bY_{(g)}^{(\mathrm{ma})} = \frac{1}{\sqrt{h_3}} \sum_{j = g}^{g + h_3 - 1} \bX_{(j)} \text{ for } g = 1, \ldots, p - h_3 + 1
$$
and $\bY^{(\mathrm{ma})} = (\bY_{(1)}^{(\mathrm{ma})}, \ldots, \bY_{(p-h_3+1)}^{(\mathrm{ma})})$ be the column-wise moving average matrix of $\bX$, where $h_3$ is a tuning parameter. We estimate the cluster labels $\bell$ by 
\be \label{eq:ma-pca}
\hat{\bell}_{\mathrm{ma}} = \operatorname{sign}(\hat{\bxi}_{\mathrm{ma}}),
\ee
where $\hat{\bxi}_{\mathrm{ma}}$ is the first eigenvector of the matrix $\bY^{(\mathrm{ma})} \bY^{(\mathrm{ma})^{\T}}$. We call this method as moving average PCA (MA-PCA) for clustering.
In the SM, we demonstrate that the signal-to-noise ratio in $\bY^{(\mathrm{ma})} \bY^{(\mathrm{ma})^{\T}}$ increases with $h_3$ when $h_3 \lesssim b$, but decreases with further increases in $h_3$ when $b = o(h_3)$.

The following theorem establishes the consistency of $\hat{\bell}_{\mathrm{ma}}$ to $\bell$ up to a sign flip.

\begin{theorem} \label{th:UB_comp_dense_cluster} 
For the clustering problem under the model in \eqref{eq:mixgauss}, under Assumptions \ref{assu:sub_gaussian}-\ref{assu:separable}, if $h_3 \asymp b$, and for some constant $c > 0$, $\bmu$ satisfies the signal-to-noise ratio condition 
\be \label{eq:signal_to_noise_MA}
\min \{ \sqrt{{n b}/{p}}, 1 \} {\|\bmu\|^2}/{\|\bSigma\|_2} > c \log n,
\ee
we have $\min\{\|\hat{\bell}_{\mathrm{ma}} + \bell\|_{\infty}, \|\hat{\bell}_{\mathrm{ma}} - \bell\|_{\infty}\} \to 0$ with probability approaching to $1$ as $n, p \to \infty$. 
\end{theorem}

Theorem \ref{th:UB_comp_dense_cluster} demonstrates that the proposed MA-PCA method can consistently cluster the observations under general covariances if the signal-to-noise ratio satisfies \eqref{eq:signal_to_noise_MA}. 
This condition aligns with the results of \cite{Abbe2022} for non-block signals where $b = 1$. In the block signal setting, the condition on the signal strength $\|\bmu\|^2$ can be weakened by a factor of $\sqrt{b}$ compared to that for non-block signals. Compared to the signal-to-noise ratio condition for CFA-PCA in Theorem \ref{th:UB_comp_sparse_cluster}, if $\|\bSigma\|_2 \le C$ and $\log p \asymp \log n$, the conditions in \eqref{eq:signal_to_noise_pre} and  \eqref{eq:signal_to_noise_MA} can be simplified as $\min\{\sqrt{n}m^{-1}, 1\} \|\bmu\|^2 > c \log p$ and $\min \{ \sqrt{n} (p / b)^{-1/2}, 1 \} \|\bmu\|^2 > c \log p$ for CFA-PCA and MA-PCA, respectively. 
When $m \ll \sqrt{p/b}$ meaning the sparse block signal regime, CFA-PCA requires a weaker condition on $\|\bmu\|$,
whereas for the dense block signal regime that $m \gg \sqrt{p/b}$, MA-PCA requires a weaker condition on $\|\bmu\|$.
This indicates the necessity of designing different algorithms for different sparsity regimes. 
The results in Section \ref{sec:lowerbound} show that both methods achieve computational minimax optimality in their respective sparsity regimes. 
A data-driven procedure to select the tuning parameter $h_3$ is provided in the simulation section.


Once the estimated cluster labels $\hat{\bell}$ are obtained, we conduct signal recovery on the sequence $\hat{\bY}_i = \hat{\ell}_i {\bX}_i$ for $i = 1, \ldots, n$, where $\hat{\bY}_i = (\hat{Y}_{i1}, \ldots, \hat{Y}_{ip})^{\T}$. 
Let $\bY_i = \ell_i \bX_i$, $\bY_0 = n^{-1/2} \sum_{i=1}^n {\bY}_i$ and
$\hat{\bY}_0 = n^{-1/2} \sum_{i=1}^n \hat{\bY}_i$, where $\hat{\bY}_0 = (\hat{Y}_{01}, \ldots, \hat{Y}_{0p})^{\T}$. If $\hat{\bell}$ converges to $\bell$, $\hat{\bY}_0$ would be close to $\bY_0$, thereby enabling signal recovery. 
For a given block $I$, let
$\hat{Y}_i(I) = |I|^{-1/2} \sum_{j \in I} \hat{Y}_{ij}$ and $\hat{Y}_0(I) =  |I|^{-1/2} \sum_{j \in I} \hat{Y}_{0j}$
be the aggregated feature over $I$. 
Let $Y_0(I) = \sum_{j \in I} Y_{0j} / \sqrt{|I|}$ and $\sigma^2(I) = {\V} \{ Y_0(I)\}$. We estimate $\sigma^2(I)$ by the pooled sample variance $\hat{\sigma}^2(I)$ of the estimated groups $\{\hat{Y}_i(I): \hat{\ell}_i = 1\}$ and $\{\hat{Y}_i(I): \hat{\ell}_i = -1\}$.

Recall that $\mathbb{J}_p(h_1) = \{(j, \ldots, j + h): 1 \leq h \leq h_1, 1 \leq j \leq p - h\}$. Take the threshold $t_p = \sqrt{4 \log(p h_1)}$. By the moderate deviation result, if $\bmu = \bzero$, we have $\mathbb{P} ( \cup_{I \in \mathbb{J}_p(h_1)} \{|\hat{Y}_0(I)| / \hat{\sigma}(I) > t_p(I)\} ) \to 0$ as $n, p \to \infty$. 
Thus, we identify the blocks with nonzero means by $\mathbb{Q}^{(1)} = \{ I \in \mathbb{J}_p(h_1) : |\hat{Y}_0(I)| / \hat{\sigma}(I) > t_p \}$.
Similar to the CFA procedure for signal identification based on $|W_0(I)|$ in Algorithm \ref{alg:2}, we use an iterative step-down procedure based on $|\hat{Y}_0(I)|$ to identify the true signal blocks and remove the overlapping blocks from $\mathbb{Q}^{(1)}$. Algorithm \ref{alg:1} outlines the proposed post-clustering signal recovery procedure. 

\begin{algorithm}\label{alg:1}
	\caption{Post-clustering block identification}
	\label{alg:Post-hoc-sgement}
	\KwIn{observation $\bX_1, \bX_2, \ldots, \bX_n$; clustering results $\hat{\bell}$; window size $h_1$. }
	\KwOut{block identification $\hat{\mathbb{I}}_{\mathrm{ma}} = \{\hat{I}_1, \ldots, \hat{I}_{\hat{m}}\}$. }  
	\BlankLine

    $\hat{\bY}_0 = n^{-1/2} \sum_{i=1}^n{\hat{\bY}_i}$, where $\hat{\bY}_i = \bX_i\hat{\ell}_i$;
    
    $k = 1$ and
    $\mathbb{Q}^{(1)} = \big\{ I \in \mathbb{J}_p(h_1) : |\hat{Y}_0(I)| / \hat{\sigma}(I) > t_p \big\}$;
    
	\While{${\mathbb{Q}}^{(k)} \neq \emptyset$}{
        $\hat{I}_k =   {\operatorname{argmax}}_{I\in {\mathbb{Q}}^{(k)}}\,\, |\hat{Y}_0(I)|$; $\mathbb{Q}^{(k+1)} = \mathbb{Q}^{(k)} \backslash \{ I\in \mathbb{Q}^{(k)}: I \cap \mathcal{E}_p(\hat{I}_k, \lfloor h_1/2 \rfloor)\ne \emptyset\}$; 
	}
    Obtain $\hat{\mathbb{I}}_{\mathrm{ma}} = \{\hat{I}_1, \hat{I}_2, \ldots, \hat{I}_{\hat{m}}\}$. 
\end{algorithm}

We use $\hat{\mathbb{I}}_{\mathrm{ma}} = \{\hat{I}_1, \hat{I}_2, \ldots, \hat{I}_{\hat{m}}\}$ to denote the $\hat{m}$ identified blocks from Algorithm \ref{alg:1}. The following theorem shows that the proposed post-clustering signal recovery procedure can consistently identify all true signal blocks. 

\begin{theorem} \label{th:UB_comp_dense_signal}
    Under the model \eqref{eq:mixgauss}, Assumptions \ref{assu:sub_gaussian}-\ref{assu:strength}, $\max \{ \| \bSigma \|_2, \| \bSigma^{-1} \|_2\} \le C$, $\sqrt{nb} \mu_0 > c \sqrt{\log p}$ for a constant $c > 0$, and additionally $\mu_j = \tau_g$ for $j \in I_{0,g}$ and $g = 1, \ldots, m$, 
    given a clustering result $\hat{\bell}$ satisfying $\mathbb{P}(\hat{\bell} = \pm \bell)= 1 - o(1)$, if $\max_g|I_{0,g}| < h_1 < d_0$, 
    we have $\mathbb{P}(\hat{m} = m) \to 1$ and 
    $\mathbb{P}( \operatorname{max}_{I \in \mathbb{I}_0} \operatorname{min}_{\hat{I} \in \hat{\mathbb{I}}_{\mathrm{ma}}} D(\hat{I}, I) > \zeta ) \to 0$ for any $\zeta > 0$. 
\end{theorem}

Similar as Theorem \ref{th:UB_comp_sparse_signal}, we also need equal signal strength in each block for consistent post-clustering signal recovery. 
The condition $\sqrt{nb} \mu_0 > c \sqrt{\log p}$ on signal strength is standard, requiring that the aggregate signal within each block goes to infinity. 
Algorithm \ref{alg:1} extends the method in \cite{Jeng2010} for IID Gaussian data with the isotropic covariance matrix to sub-Gaussian data with unknown clusters. Note that the MA-PCA and associated signal recovery procedures for dense block signals work under general covariances, which don't require the bandable covariances assumption in \eqref{eq:bandable}.


\section{Minimax lower bounds} \label{sec:lowerbound}

In this section, we establish the statistical and computational minimax lower bounds (SMLB and CMLB) for clustering and signal recovery, and show the proposed CFA-PCA and MA-PCA procedures are computationally minimax optimal. 
Algorithms can be categorized according to computational feasibility. For example, exact $k$-means and subset selection are NP-hard, which cannot be solved in polynomial time of sample size and dimension. In contrast, methods based on PCA are computationally feasible. The SMLBs provide the regions of signals where any clustering or signal recovery procedure would fail. While, the computational minimax results are restricted to polynomial time methods. 

Let \(\lfloor \cdot \rfloor\) denote the floor function. 
In order to derive the sharp minimax boundaries, we consider the standard multivariate Gaussian distribution $F = N(\bzero, \bI_p)$ for the data in this section and the mean parameter $\bmu$ from the class
\begin{equation} \label{eq:signal_class}
\begin{aligned}
\mathcal{U}(\alpha, \beta, r) = \big\{ \bmu :\ & \text{there exist } m = \lfloor p^{1 - \alpha - \beta} \rfloor \text{ blocks with } |I_{0,g}| = \lfloor \rho_g p^{\alpha} \rfloor, \\ 
& \mu_j = \tau_g \text{ for } j \in I_{0,g} \text{ and } |\tau_g| = \tau = p^{-r}, \mu_j = 0 \text{ otherwise}, \\
& \text{and } d_0 > \rho_0 \log(p) p^{\alpha} \big\},
\end{aligned}
\end{equation}
where \(\alpha\), \(\beta\) and $r$ are the block size, signal sparsity and signal strength parameters, respectively, satisfying $\alpha, \beta, r \in (0, 1)$, \(0 < \alpha + \beta \leq 1\) and \(\rho_0, \rho_1, \ldots, \rho_m > 0\). Here, we impose the specific parameterization $m = \lfloor p^{1 - \alpha - \beta} \rfloor$, $b = p^{\alpha}$ and $|\mu_j| = p^{-r}$ for all $j \in S(\bmu)$ to the general setting in Section \ref{sec:settings}. We focus on the high-dimensional setting $n = p^{\theta}$ for $0< \theta < 1 - \alpha$, namely, $n = o(p^{1-\alpha})$, where $p^{1-\alpha}$ is the dimension of the aggregated variables by the blocks of size $p^{\alpha}$. 
Under the class $\mathcal{U}(\alpha, \beta, r)$, 
the sparse and dense block signal regimes correspond to \(\beta \in ((1-\alpha)/2, 1 - \alpha)\) and \(\beta \in (0, (1-\alpha)/2)\), respectively.
The signal structure under $\mathcal{U}(\alpha, \beta, r)$ is similar to the setting of block signals for hypothesis testing in \cite{Kou2022}, and follows the parameterization of signal sparsity and strength in \cite{Donoho2004, Jin2017}. 

For the estimated labels \(\hat{\bell} = (\hat{\ell}_1, \ldots, \hat{\ell}_n)^{\T}\) from a clustering procedure, we measure its performance using the expected Hamming distance, defined as
\begin{equation} \label{eq:hamm_cluster}
    \operatorname{Hamm}_{\mathrm{clu}}(\hat{\bell}; \bmu, \varpi) = \inf_{\pi = \pm 1} n^{-1} \mathbb{E}| \{i: \hat{\ell}_i \ne \pi \ell_i \} |,
\end{equation} 
where the expectation is taken with respect to the class labels $\bell$ and the random noise $\bZ$ in \eqref{eq:mixgauss}. 
Similarly, for the estimated signal set \(\hat{S}\) from a signal recovery procedure, we measure its performance using the normalized expected Hamming distance, defined as
\begin{equation} \label{eq:hamm_signal}
    \operatorname{Hamm}_{\mathrm{sig}}(\hat{S}; \bmu, \varpi) = (p \varepsilon)^{-1}   \mathbb{E} \big| \hat{S} \Delta S(\bmu) \big|,
\end{equation}
where \(S(\bmu)\) is the true signal set of $\bmu$, $\varepsilon = |S(\bmu)| / p = s / p$ is the proportion of signals out of $p$ variables, and \(A \Delta B = (A \setminus B) \cup (B \setminus A)\) denotes the difference between two sets.


We use the low-degree polynomial conjecture \citep{Hopkins2018} to establish our computational minimax lower bounds (CMLBs). This conjecture has been widely applied to various problems \citep{Kunisky2022, Schramm2022}. The CMLBs for clustering and signal recovery can be developed based on the CMLB for the hypotheses testing problem 
\begin{equation} \label{eq:nullhyp}
    H_0^{(p)}: \bX_i \stackrel{\text{IID}}{\sim} N(\bzero, \mathbf{I}_p) 
    \mbox{ \ vs. \ } 
    H_1^{(p)}: \bX_i \stackrel{\text{IID}}{\sim} \frac{\varpi}{1+\varpi} N(\bmu, \mathbf{I}_p) + \frac{1}{1+\varpi} N(-\varpi \bmu, \mathbf{I}_p).
\end{equation}
\cite{Loffler2022, Lei2023} used similar approaches of deriving computational lower bounds through hypotheses testing for clustering under sparse non-block signals and network analysis, respectively. 
Let \( \mathbb{P}_0 \) and \( \mathbb{P}_1 \) be the joint distribution of the data \( \{\bX_i\} \) under the null and alternative hypotheses in \eqref{eq:nullhyp}, respectively. 
Let \(\Lambda(\bX) = {\mathrm{d} \mathbb{P}_1(\bX)} / {\mathrm{d} \mathbb{P}_0(\bX)}\) be the likelihood ratio, \(\Lambda_{(D)}\) be the orthogonal projection of \(\Lambda(\bX)\) onto the linear subspace of polynomials of degree at most \(D\), and $\mathbb{E}_0( \cdot )$ be the expectation under the null distribution \(\mathbb{P}_0\).
The low-degree polynomial conjecture is stated as follows.

\begin{conjecture}[Low-degree polynomial conjecture]\label{conj:lowdegree}
If \( \mathbb{E}_0\{ \Lambda_{(D)}^2 \} = 1 + o(1)\), with \(D = \log^{1.01}(p)\), then \( \mathbb{P}_0 \) and \( \mathbb{P}_1 \) are indistinguishable by any polynomial-time algorithm.
\end{conjecture}

Let $c_1(\theta, \alpha) = (1 - \theta - \alpha) / 2$ and $c_2(\theta, \alpha) = 1 - \theta / 2 - \alpha$.
For \(\theta, \alpha \in (0,1)\), we introduce the functions \(\eta_{\theta, \alpha}^{\mathrm{clu}}(\beta)\) and \(\eta_{\theta, \alpha}^{\mathrm{sig}}(\beta)\) for the statistical minimax lower bounds, where
\be\nonumber
\begin{split}
\eta _{\theta, \alpha}^{\mathrm{clu}}\left( \beta \right) &=
\{(1 + \theta + \alpha - 2\beta) / 4\} \mathbbm{1}\{\beta < c_1(\theta, \alpha)\} 
\\
&+ \{(\theta + \alpha) / 2\} \mathbbm{1}\{c_1(\theta, \alpha) < \beta < 2c_1(\theta, \alpha)\} + \{(1 - \beta) / 2\} \mathbbm{1}\{2c_1(\theta, \alpha) < \beta < 1 - \alpha\}, \\
\eta _{\theta, \alpha}^{\mathrm{sig}}\left( \beta \right) &=
\{(\theta + \alpha) / 2\} \mathbbm{1}\{\beta < 2c_1(\theta, \alpha)\} + \{(1 + \theta + \alpha - \beta) / 4\} \mathbbm{1}\{2c_1(\theta, \alpha) < \beta < 1 - \alpha\}.
\end{split} 
\ee
The CMLB $\tilde{\eta}_{\theta, \alpha}^{\mathrm{hyp}}( \beta )$ for testing the hypotheses in \eqref{eq:nullhyp} is derived in the SM using Conjecture \ref{conj:lowdegree}, where $\tilde{\eta}_{\theta, \alpha}^{\mathrm{hyp}}( \beta )$ is defined in \eqref{sm-eq:hyp_clu_lowerbound} in the SM. It is shown in Lemma \ref{lemma:clu_sig_hyp} in the SM that the CMLBs for clustering and signal recovery are no larger than that for testing \eqref{eq:nullhyp}. Meanwhile, 
a CMLB is no larger than the corresponding SMLB. Therefore, we take
\(\tilde{\eta}_{\theta, \alpha}^{\mathrm{clu}}(\beta) = \min\{\eta_{\theta, \alpha}^{\mathrm{clu}}( \beta ), \tilde{\eta}_{\theta, \alpha}^{\mathrm{hyp}}( \beta )\} \) and \(\tilde{\eta}_{\theta, \alpha}^{\mathrm{sig}}(\beta) = \min\{\eta_{\theta, \alpha}^{\mathrm{sig}}( \beta ), \tilde{\eta}_{\theta, \alpha}^{\mathrm{hyp}}( \beta )\}\) for our computational minimax results, 
where 
\begin{equation} \nonumber
\begin{split}
\tilde{\eta} _{\theta, \alpha}^{\mathrm{clu}}( \beta ) & = 
\{(1 + \theta + \alpha - 2\beta) / 4\} \mathbbm{1}\{\beta < (1 - \alpha) / 2\} \\
& + (\theta / 4 + \alpha / 2) \mathbbm{1}\{ (1 - \alpha) / 2 < \beta < c_2(\theta, \alpha)\} + \{(1 - \beta) / 2\} \mathbbm{1}\{ c_2(\theta, \alpha) < \beta < 1 - \alpha\}, \\
\tilde{\eta}_{\theta, \alpha}^{\mathrm{sig}}( \beta ) &=
\{(\theta + \alpha) / 2\} \mathbbm{1}\{\beta < c_1(\theta, \alpha)\} + \{(1 + \theta + \alpha - 2\beta) / 4\} \mathbbm{1}\{c_1(\theta, \alpha) < \beta < (1 - \alpha) / 2\} \\
& + (\theta / 4 + \alpha / 2) \mathbbm{1}\{ (1 - \alpha) / 2 < \beta < 1 - \alpha\}.
\end{split}
\end{equation}

The following two theorems state the statistical and computational minimax lower bounds for clustering and signal recovery under the block signal setting. 

\begin{theorem} \label{th:clu_LB}
Under the model in \eqref{eq:mixgauss} with $\bZ_i \sim N(\bzero, \bI_p)$,
given \(\alpha, \theta \in (0,1)\), $\theta < 1 - \alpha$, \(\beta \in (0, 1 - \alpha)\) and $\varpi > 0$, we have 
$\lim_{p \rightarrow \infty} \, \sup_{\bmu \in \mathcal{U}(\alpha, \beta, r)} \operatorname{Hamm}_{\mathrm{clu}}( \hat{\bell}; \bmu, \varpi) \geq {\min \{ 1, \varpi \} } / (1+\varpi)$ for (1) any clustering procedure \(\hat{\bell}\) if \(r > \eta_{\theta, \alpha}^{\mathrm{clu}}(\beta)\) and (2) any polynomial-time clustering procedure \(\hat{\bell}\) if \(r > \tilde{\eta}_{\theta, \alpha}^{\mathrm{clu}}(\beta)\).
\end{theorem}

\begin{theorem}\label{th:sig_LB}
Under the model in \eqref{eq:mixgauss} with $\bZ_i \sim N(\bzero, \bI_p)$,
given \(\alpha, \theta \in (0,1)\), $\theta < 1 - \alpha$, \(\beta \in (0, 1 - \alpha)\) and $\varpi > 0$, 
we have 
$\lim_{p \rightarrow \infty} \, \sup_{\bmu \in \mathcal{U}(\alpha, \beta, r)} \operatorname{Hamm}_{\mathrm{sig}}( \hat{S}; \bmu, \varpi) \geq C > 0$
for (1) any signal recovery procedure \(\hat{S}\) if \(r > \eta_{\theta, \alpha}^{\mathrm{sig}}(\beta)\) and (2) any polynomial-time signal recovery procedure \(\hat{S}\) if \(r > \tilde{\eta}_{\theta, \alpha}^{\mathrm{sig}}(\beta)\). 
\end{theorem}

Theorems \ref{th:clu_LB} and \ref{th:sig_LB} characterize both the statistical and computational minimax lower bounds for clustering and signal recovery under block signals. They demonstrate that when $r > \eta_{\theta, \alpha}^{\mathrm{clu}}(\beta)$ and $r > \eta_{\theta, \alpha}^{\mathrm{sig}}(\beta)$, no statistical algorithm can achieve consistent clustering and signal recovery, respectively. Note that the signal strength $\tau$ decreases with the increase of $r$ 
in the class $\mathcal{U}(\alpha, \beta, r)$, and the error rate ${\min \{ 1, \varpi \} }/({1+\varpi})$ can be achieved by simply guessing $\hat{\ell}_i = 1$ or $\hat{\ell}_i = -1$ for all $i$. 
In particular, when $\alpha = 0$, representing signals without block structures, the derived SMLBs align with those in \cite{Jin2017} under a Bayesian setting of randomly located signals.  
The two theorems also show that $\tilde{\eta}_{\theta, \alpha}^{\mathrm{clu}}(\beta)$ and $\tilde{\eta}_{\theta, \alpha}^{\mathrm{sig}}(\beta)$ are the CMLBs for clustering and signal recovery. Namely, no polynomial-time algorithm can achieve consistent clustering and signal recovery when $r > \tilde{\eta}_{\theta, \alpha}^{\mathrm{clu}}(\beta)$ and $r > \tilde{\eta}_{\theta, \alpha}^{\mathrm{sig}}(\beta)$, respectively.
This delineates a ``region of impossibility'' and provides a lower bound on the necessary signal strength for successful clustering and signal recovery under block signals and polynomial-time computational constraint. Meanwhile, note that the four functions \(\eta_{\theta, \alpha}^{\mathrm{clu}}(\beta)\), \(\eta_{\theta, \alpha}^{\mathrm{sig}}(\beta)\), \(\tilde{\eta}_{\theta, \alpha}^{\mathrm{clu}}(\beta)\) and \(\tilde{\eta}_{\theta, \alpha}^{\mathrm{sig}}(\beta)\) increase as $\theta$ or $\alpha$ increases, which means that the impossible regions shrink with increasing sample size and length of signal blocks. 

\begin{figure}[h]
    \centering
    \includegraphics[width = \textwidth]{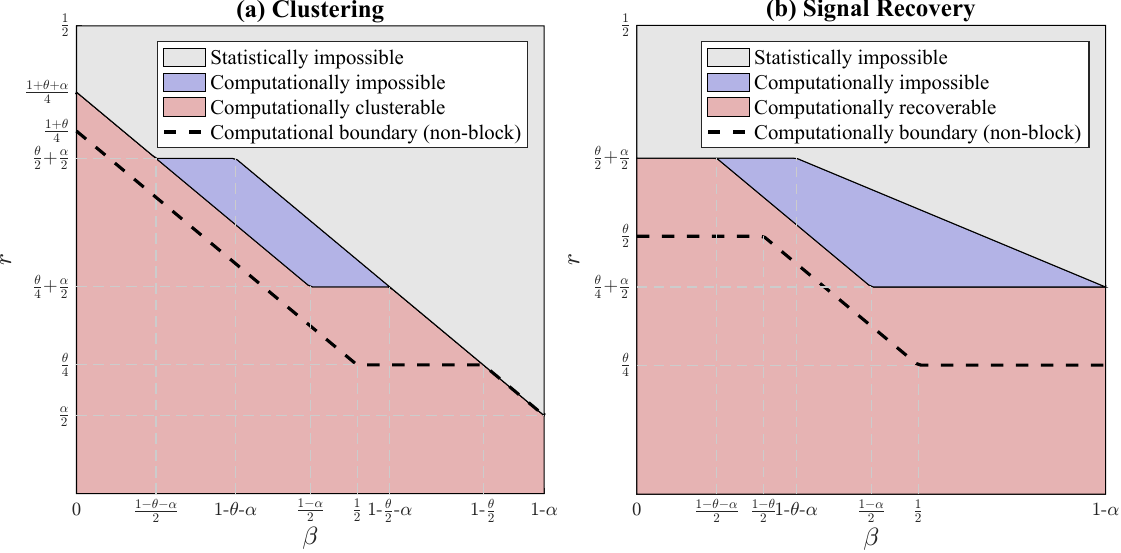}
    \caption{Phase transition of clustering in panel (a) and signal recovery in panel (b). Gray region: consistent clustering or signal recovery is impossible by any algorithm. Purple region: consistent clustering or signal recovery is impossible under the polynomial computation time constraint. Red region: consistent clustering or signal recovery can be achieved by a polynomial-time algorithm. Black dash line: CMLBs under non-block signals ($\alpha = 0$). 
    }
    \label{fig:phase}
\end{figure}

Panels (a) and (b) in Figure \ref{fig:phase} illustrate the SMLBs and CMLBs. From those plots, there is a statistical-computational gap that \(\tilde{\eta}_{\theta, \alpha}^{\mathrm{clu}} < {\eta}_{\theta, \alpha}^{\mathrm{clu}}\) for clustering in the moderately sparse regime of \( ({1 - \theta - \alpha})/{2} < \beta < 1 - {\theta}/{2} - \alpha \). For signal recovery, a statistical-computational gap exists except the regime of \(\beta < (1 - \theta - \alpha)/{2} \). 
Figure \ref{fig:phase} also shows 
the computationally impossible regions are much smaller under block signals, which demonstrates the benefit in the minimax lower bounds gained by considering the block structure of signals. 
%

It is worth noting that \cite{Loffler2022} derived the computational minimax threshold for the signal-to-noise ratio in the high-dimensional clustering problem, which corresponds to the special case of $\alpha = 0$ in our setting for non-block signals. 
However, they didn't establish the phase transition diagram, nor study the minimax results for signal recovery. 
Therefore, their results don't show the statistical-computational gaps in the phase transition diagram, and cannot compare the difficulty of clustering and signal recovery across different sparsity regimes from a minimax view. 

Let $\hat{S}_{\mathrm{cfa}} = \cup_{\hat{I} \in \hat{\mathbb{I}}_{\mathrm{cfa}}} \hat{I}$ and $\hat{S}_{\mathrm{ma}} = \cup_{\hat{I} \in \hat{\mathbb{I}}_{\mathrm{ma}}} \hat{I}$ be the sets of identified signals by Algorithms \ref{alg:2} and \ref{alg:1}, respectively.
From the results of Theorems \ref{th:UB_comp_sparse_cluster}--\ref{th:UB_comp_dense_signal}, the following corollary shows that the proposed CFA-PCA procedure ($\hat{\bell}_{\mathrm{cfa}}$ and $\hat{S}_{\mathrm{cfa}}$) and the MA-PCA procedure ($\hat{\bell}_{\mathrm{ma}}$ and $\hat{S}_{\mathrm{ma}}$) can 
achieve the CMLBs for clustering and signal recovery under the Gaussian distribution with identity covariance in the sparse and dense block signal regimes, respectively. 

\begin{corollary} \label{coro:opt_clu_sig}
Under the model in \eqref{eq:mixgauss} with $\bZ_i \sim N(\bzero, \bI_p)$, given \(\alpha, \theta \in (0,1)\), $\theta < 1 - \alpha$, \(\beta \in (0, 1 - \alpha)\) and $\varpi > 0$, choosing $\max_g |I_{0,g}| < h_1 < d_0$, $h_1 \le h_2 < \tilde{d}$ and $h_3 \asymp p^{\alpha}$,
then 
\be\nonumber
\begin{split}
& \lim_{n, p \rightarrow \infty} \, \sup_{\bmu \in \mathcal{U}(\alpha, \beta, r)} \operatorname{Hamm}_{\mathrm{clu}}( \hat{\bell}; \bmu, \varpi) \rightarrow 0 \mbox{ for $r < \tilde{\eta}_{\theta, \alpha}^{\mathrm{clu}}(\beta)$ and } \\
& \lim_{n, p \rightarrow \infty} \, \sup_{\bmu \in \mathcal{U}(\alpha, \beta, r)} \operatorname{Hamm}_{\mathrm{sig}}( \hat{S}; \bmu, \varpi) \rightarrow 0 \mbox{ for $r < \tilde{\eta}_{\theta, \alpha}^{\mathrm{sig}}(\beta)$},
\end{split}
\ee
where $\hat{\bell} = \hat{\bell}_{\mathrm{ma}}$ and $\hat{S} = \hat{S}_{\mathrm{ma}}$ if $\beta < (1-\alpha) / 2$, and $\hat{\bell} = \hat{\bell}_{\mathrm{cfa}}$ and $\hat{S} = \hat{S}_{\mathrm{cfa}}$ if $\beta > (1-\alpha) / 2$. 
\end{corollary}

Corollary \ref{coro:opt_clu_sig} shows that the proposed  CFA-PCA and MA-PCA procedures are computationally minimax optimal in the sparse and dense block signal regimes, respectively. 
Also see Table \ref{tab:optimal} in the SM for a summary of the optimality results.
It also demonstrates the CMLBs derived in Theorems \ref{th:clu_LB} and \ref{th:sig_LB} are tight. 

\section{Extensions to tensor data} \label{sec:ext}

This section extends the proposed methods to matrix- and tensor-valued data, commonly observed in geoscience and medical imaging studies. 
Suppose we observe $n$ independent order-$q$ tensors with dimensions $p_1, \ldots, p_q$, denoted by $\{\mathcal{X}_i = (X_{i, j_1 \ldots j_q}) \}_{i=1}^n$, where $j_t$ denotes the $j_t$th variable observed in the $t$th mode of the tensor for $1 \leq j_t \leq p_t$. Let $p = \prod_{t=1}^q p_t$ be the total dimension of the tensor. 
Extending \eqref{eq:mixgauss} to tensor data, we assume $\{\mathcal{X}_i\}$ satisfy the cluster model
$\mathcal{X}_i = {\tilde{\ell}}_i \mathscr{U} + \mathcal{Z}_i$, for $i = 1, 2, \ldots, n$,
where $\mathscr{U} \in \mathbb{R}^{p_1 \times \ldots \times p_q}$ is the mean of the data representing the signals for clustering, and $\mathcal{Z}_i \in \mathbb{R}^{p_1 \times \ldots \times p_q}$ represents random noise. Our primary goal is to estimate the cluster labels $\bell = (\ell_1, \ldots, \ell_n)^{\T}$ and recover the signal set $\mathcal{S}(\mathscr{U}) = \{(j_1, j_2, \ldots, j_q) : \mathscr{U}_{j_1, j_2, \ldots, j_q} \neq 0 \}$, where $\ell_i = \operatorname{sign}(\tilde{\ell}_i)$. Let $\mathcal{I}_{0,1}, \ldots, \mathcal{I}_{0,m}$ denote the $m$ non-overlapping tensor blocks of nonzero means, where $\mathcal{I}_{0,g} = I_{0,g}^{(1)} \times \cdots \times I_{0,g}^{(q)}$ and $I_{0,g}^{(t)} = \{j_g^{(t)}, \ldots, j_g^{(t)} + l_g^{(t)}\}$ 
is a block in the $t$th mode for $t = 1, \ldots, q$ and $g = 1, \ldots, m$. 
%
Theorem \ref{th:tensor_LB} in SM shows that the SMLBs and CMLBs derived for vector-valued data can be extended to tensor-valued data. In the following, we extend the proposed MA-PCA and CFA-PCA procedures to tensor-valued data. 

For the CFA-PCA method, 
let $\mathcal{J}_{p}(h_1) = \mathbb{J}_{p_1}(h_1) \times \cdots \times \mathbb{J}_{p_q}(h_1)$, where $\mathbb{J}_{p_t}(h_1) = \{ \{j, \ldots, j+h \}: 1\le h \le h_1, 1 \le j < j+h \le p_t \}$.
For any pair of blocks $\mathcal{I}_1, \mathcal{I}_2 \in \mathcal{J}(h_1)$, let $X_i(\mathcal{I}_1) = \sum_{(j_1, \ldots, j_q) \in \mathcal{I}_1} X_{i, j_1 \ldots j_q} / \sqrt{|\mathcal{I}_1|}$ and $ W_i(\mathcal{I}_1, \mathcal{I}_2) = X_i(\mathcal{I}_1) X_i(\mathcal{I}_2)$.
Similar as \eqref{eq:pair_segment}, for each $\mathcal{I}_1 \in \mathcal{J}(h_1)$, we can define the maximal cross-block product $W_0(\mathcal{I}_1)$ based on $W_i(\mathcal{I}_1, \mathcal{I}_2)$. 
Let $\mathscr{E}(\mathcal{I}, \tilde{c}) = \mathcal{E}_{p_1}( I^{(1)},  \tilde{c}) \times \cdots \times \mathcal{E}_{p_q}( I^{(q)},  \tilde{c})$ be the expanding operator for tensor blocks, 
where $\tilde{c}$ is the expanded window length and $\mathcal{E}_{p}(I^{(t)},  \tilde{c})$ is the expanded block of $I^{(t)}$ defined in \eqref{eq:expand_operator}. Using $W_0(\mathcal{I})$, the CFA method in Algorithm \ref{alg:2} can be extended to tensor-valued data to identify the tensor blocks with nonzero means. 
Once the estimated blocks \(\hat{\mathcal{I}}_1, \ldots, \hat{\mathcal{I}}_{\hat{m}}\) are obtained from Algorithm \ref{alg:2}, we construct the aggregated data matrix for those selected blocks as 
$\hat{\bY}^{\mathrm{(cfa)}} = (\hat{\bY}^{\mathrm{(cfa)}}_{(1)}, \ldots, \hat{\bY}^{\mathrm{(cfa)}}_{(\hat{m})})$, where $\hat{\bY}^{\mathrm{(cfa)}}_{(k)} = |\hat{\mathcal{I}}_k|^{-1/2} {\sum}_{(j_1, \ldots, j_q) \in \hat{\mathcal{I}}_k} \bX_{(j_1 \ldots j_q)}$ for $k = 1, \ldots, \hat{m}$. 
The cluster labels $\bell$ can be estimated by $\hat{\bell}_{\mathrm{cfa}} = \operatorname{sign}(\hat{\bxi}_{\mathrm{cfa}})$, where $\hat{\bxi}_{\mathrm{cfa}}$ is the first eigenvector of the matrix $\hat{\bY}^{(\mathrm{cfa})} \hat{\bY}^{(\mathrm{cfa}) ^{\T}}$. 

For the tensor-valued MA-PCA method, let $\bX_{(j_1 \ldots j_q)} = (X_{1, j_1 \ldots j_q}, \ldots, X_{n, j_1 \ldots j_q} )^{\T}$ and 
$\bY^{\mathrm{(ma)}}_{(g_1 \ldots g_q)} = \frac{1}{\sqrt{h_3^{q}}} \sum_{j_1 = g_1}^{g_1 + h_3 - 1} \ldots \sum_{j_q = g_q}^{g_q + h_3 - 1} \bX_{(j_1 \ldots j_q)}$ 
be the moving average with a window length $h_3$ on each mode.
Gather all such vectors $\bY^{\mathrm{(ma)}}_{(g_1 \ldots g_q)}$ into a large data matrix $\bY^{(\mathrm{ma})} \in \mathbb{R}^{n \times \tilde{p}}$, where $\tilde{p} = \prod_{t=1}^q (p_t - h_3 + 1)$.
The MA-PCA estimate of $\bell$ is given by $\hat{\bell}_{\mathrm{ma}} = \operatorname{sign}(\hat{\bxi}_{\mathrm{ma}})$, where $\hat{\bxi}_{\mathrm{ma}}$ is the first eigenvector of the matrix $\bY^{(\mathrm{ma})} \bY^{(\mathrm{ma})^{\T}}$. 
Given estimated cluster labels $\hat{\bell}$, we perform post-clustering signal recovery on the tensor $\{ \hat{\mathcal{Y}}_i = \hat{\ell}_i \mathcal{X}_i\}_{i=1}^n$. 
For a given block $\mathcal{I} = I^{(1)} \times \cdots \times I^{(q)} \in \mathcal{J}_{p}(h_1)$, we define
$\hat{Y}_i(\mathcal{I}) =  \sum_{(j_1, \ldots, j_q) \in \mathcal{I}} \hat{\mathcal{Y}}_{i, j_1 \ldots j_q} / \sqrt{ |\mathcal{I}|} $ and $\hat{Y}_0(\mathcal{I}) = \sum_{i=1}^n \hat{Y}_i(\mathcal{I}) / \sqrt{n}$, where $|\mathcal{I}| = |I^{(1)}| \cdots  | I^{(q)} |$ is the cardinality of $\mathcal{I}$. 
We apply Algorithm \ref{alg:1} on $\{\hat{Y}_i(\mathcal{I})\}_{i = 1}^{n}$ and $\hat{Y}_0(\mathcal{I})$ for $\mathcal{I} \in \mathcal{J}_{p}(h_1)$ to identity the blocks with signals.

We conjecture the tensor MA-PCA and CFA-PCA procedures can also achieve the CMLBs in Theorem \ref{th:tensor_LB} for clustering and signal recovery under the dense and sparse signal regimes, respectively. We leave the theoretical justification in future works. In Section \ref{sec:exp}, we conduct simulation experiments using matrix-valued data, demonstrating good performances of the proposed procedures under different regimes of signals. 

\section{Simulation}\label{sec:exp}

In this section, we evaluate the proposed methods MA-PCA and CFA-PCA for clustering and signal recovery, and compare them with the IF-PCA \citep{Jin2016}, spectral clustering and $k$-means by numerical experiments. Note that the latter three algorithms are not designed for block signals. We combine them with Algorithm \ref{alg:Post-hoc-sgement} for post-clustering block identification to evaluate their performance on signal recovery in our setting. 

We consider two-group clustering for matrix-valued data. 
We generated 
$\bX_i = \ell_i \bU + \bZ_i$ and $\ell_i \sim 2\mbox{Bernoulli}( 1 / 2 ) - 1$ independently for $i = 1, \ldots, n$, where $\bU = (U_{j_1j_2})$ is a $p_1 \times p_2$ signal matrix with $m$ signal blocks and other entries being $0$, and $\bZ_i = (Z_{i, j_1j_2})$ is a $p_1 \times p_2$ noise matrix with entries $Z_{i, j_1j_2} \overset{i.i.d.}{\sim} N(0, 1)$. We set $\theta = 0.4$ and the total sample size was $n = 2 \lfloor ( p_1p_2 )^{\theta}/2 \rfloor$, where $\lfloor \cdot \rfloor$ is the floor function. The number of block was $m = \lfloor ( p_1p_2 )^{1-\alpha-\beta} \rfloor$. 
For the cases of dense and sparse block signals, we set $\alpha = 0.5, \beta = 0.24$ and $\alpha = 0.3, \beta = 0.6$, respectively. The block size along each coordinate ranged from $L_{\min} = \lfloor \rho _{\min} p_{1}^{\alpha} \rfloor$ to $L_{\max} =\lfloor \rho _{\max} p_{1}^{\alpha} \rfloor$ with a minimum gap between two blocks $d_0 = \lfloor \rho _{0} p_{1}^{\alpha} \rfloor$. The parameters were set as $\rho _{\min} = 0.8$, $\rho _{\max} = 1.25$ and $\rho_0 = 1.5$. 

The locations of the signal blocks were randomly selected in the matrix $\bU$. We first randomly chose the center of the first block from the set $\{1, \ldots, p_1\}\times\{1, \ldots, p_2\}$ excluding the border band of size $L_{\max} / 2$. Then, the block sizes along the two coordinates were randomly drawn from a uniform distribution over $[L_{\min}, L_{\max}]$. After the location of the first block, denoted as $\mathcal{I}_{0,1}$, was selected, we removed the region $\mathscr{E}(\mathcal{I}_{0, 1}, d_0)$ that expands $\mathcal{I}_{0, 1}$ by $d_0$ from the candidate set of grids for selecting the next block. This guarantees the randomly selected blocks to be at least $d_0$ separated. The rest of the blocks were selected from the remaining grids in the same manner sequentially, until a total of $m$ blocks were identified. For the entries in a selected block $\mathcal{I}_{0,g}$, we assigned $U_{j_1j_2} = \tau_g$ for $(j_1, j_2) \in \mathcal{I}_{0,g}$, where $\tau_g = \tau$ or $-\tau$ with equal probability. We set the dimensions $p_1 = p_2 = 50, 100, 200$, corresponding to $p = p_1p_2 = 2500, 10000, 40000$ and $n = 22, 38, 68$. For each combination of $(\alpha, \beta)$ and $(n, p)$, we considered a sequence of values for the signal strength $\tau$. The specific settings are provided in Table \ref{tab:sim_settings} in the SM. Once $\bU$ was generated, it was kept fix through 500 repetitions for each setting.

For a fair comparison with the existing methods, IF-PCA, spectral clustering and $k$-means, which don't consider block signals, we also evaluated the proposed methods under non-block signals where the signal components were randomly selected. Under this setting, $\alpha = 0$ and we set $\beta = 0.24$ and $0.8$ for the cases of dense and sparse signals, respectively. 




We propose a data-driven method to select the window sizes. 
For MA-PCA, let $\hat{s}(h_1, h_3)$ be the number of identified signals using the hyperparameters $h_1$ and $h_3$. From the results of sensitivity analysis in the SM, we find that the smallest signal recovery loss is typically corresponds to the smallest clustering loss. Intuitively, for post-clustering block identification, the signals can be identified if the estimated cluster labels are close to the truth. This motivates us to use the number of significant variables $\hat{s}(h_1, h_3)$ as the criterion. Given a candidate set $\mathcal{H} = \{(h_1, h_3): 1\le h_1 \le h_3 \le h_{\max} \}$, let $\hat{s}_{\max} = \max_{(h_1, h_3) \in \mathcal{H}} \hat{s}(h_1, h_3)$. We select the window size $(\hat{h}_1, \hat{h}_3)$ as 
\be \label{eq:select-h} 
(\hat{h}_1, \hat{h}_3) = \mathop{\operatorname{argmin}} \limits_{(h_1, h_3)} \{ h_1:  \hat{s}(h_1, h_3) > (1-\epsilon) \hat{s}_{\max} \}, 
\ee
where $\epsilon$ is a small positive constant. 
In the simulation, we set $\epsilon = 0.01$ and $h_{\max} = 15,25,30$ for $p_1 = 50, 100, 200$, respectively.
Similarly, although the CFA signal recovery procedure is conducted before clustering, we also use the post-clustering signal recovery result to select the hyperparameters $h_1$ and $h_2$ for CFA-PCA, which is described in the SM. 
As an illustration example, Figure \ref{fig:hyper} in the SM shows that 
the hyperparameters chosen by \eqref{eq:select-h} in one repetition of the simulation, which minimized both clustering error and signal recovery error in this case. 




\begin{figure}[h]
    \centering
    \includegraphics[width = \textwidth]{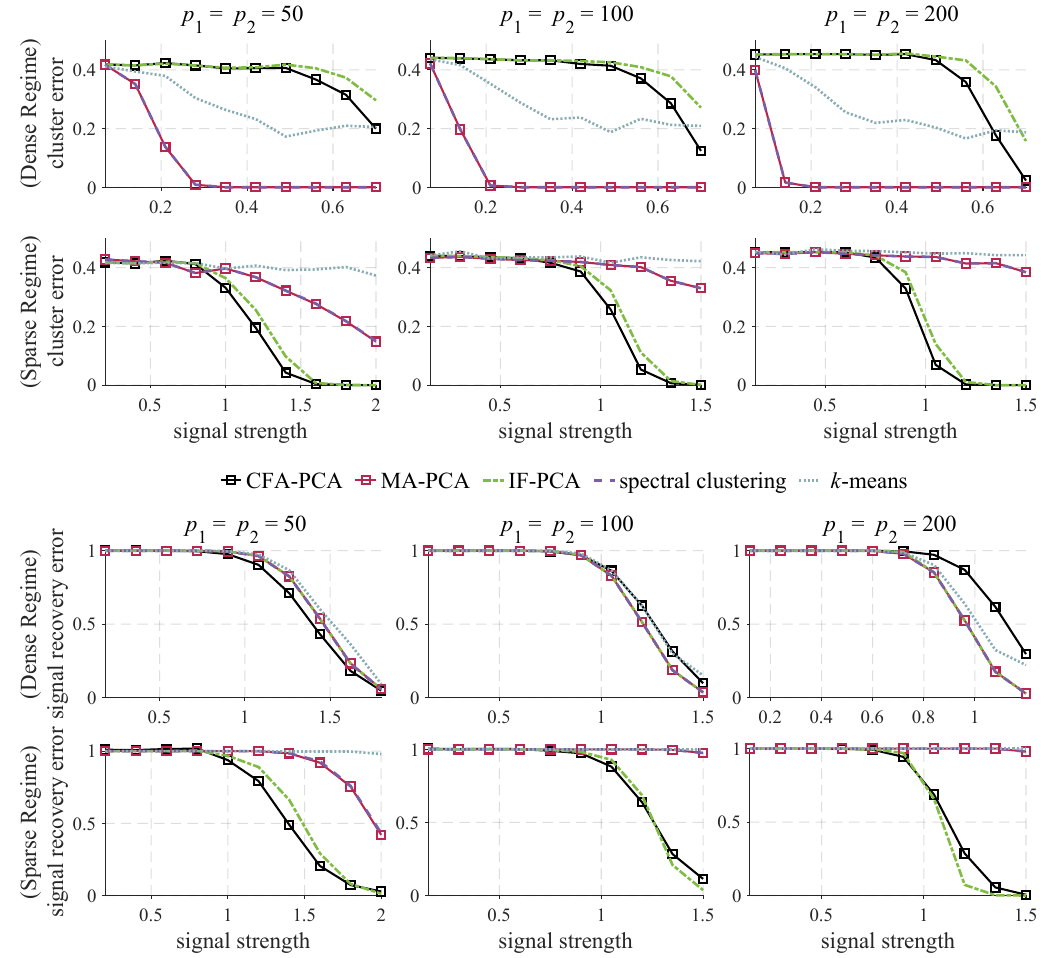}
    \caption{Simulation results under the non-block signal settings. The clustering errors (upper panels) and signal recovery errors (lower panels) of CFA-PCA, MA-PCA, IF-PCA, spectral clustering and $k$-means under different signal strengths, averaged over 500 repeated experiments. The rows represent the dense and sparse signal scenarios and the columns represent different dimensions.}
    \label{fig:exp_errors_noblock}
\end{figure}

We first present the results under the non-block setting in Figure \ref{fig:exp_errors_noblock}, where $\alpha = 0$ and $\beta = 0.24$ and $0.8$ for the dense and sparse signals, respectively. 
We chose a proper range of signal strength to reflect the change of clustering and signal recovery errors from 1 to 0 in each setting, as shown in Figure \ref{fig:exp_errors_noblock}. From this figure, we observe that the clustering errors of MA-PCA and spectral clustering were quite similar and better than CFA-PCA and IF-PCA under the dense signal setting, whereas, the clustering errors of CFA-PCA and IF-PCA were quite similar and better than MA-PCA and spectral clustering under the sparse signal setting. The performance of $k$-means was ranked in the middle for dense signals, but was the worst for sparse signals. Meanwhile, for signal recovery, CFA-PCA and IF-PCA were comparable and much better than the other three methods under the sparse signal setting. While, the performance of MA-PCA, IF-PCA and spectral clustering were similar and became the best as the increase of $p$ for the dense signal setting. Also notice that, under this setting, signal recovery requires much larger signal strength than clustering so that the loss diminished to 0. Those results demonstrate that the proposed methods don't suffer efficiency loss by considering block structures for non-block signals. Their performances were comparable to the existing methods under the non-block setting. 

\begin{figure}[h]
    \centering
    \includegraphics[width = \textwidth]{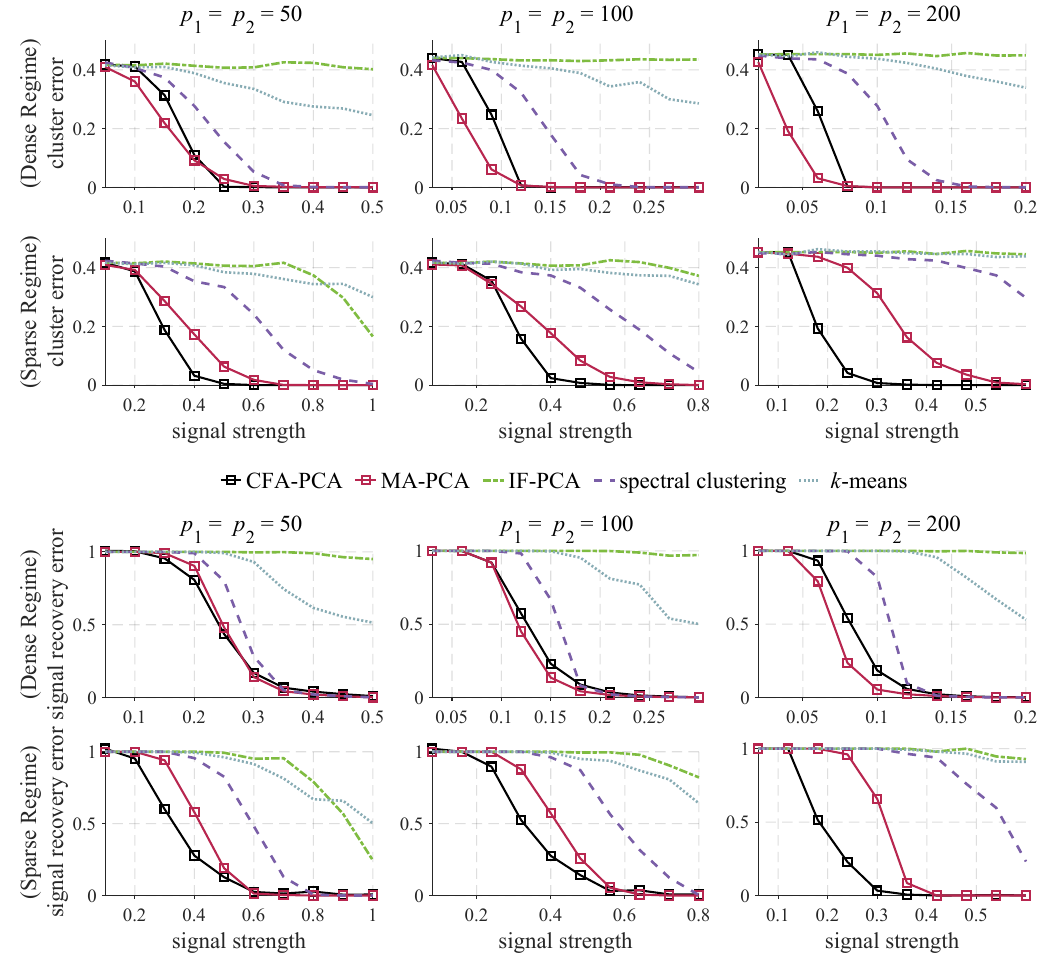}
    \caption{Simulation results under the block signal settings in Table \ref{tab:sim_settings}. The clustering errors (upper panels) and signal recovery errors (lower panels) of CFA-PCA, MA-PCA, IF-PCA, spectral clustering and $k$-means under different signal strengths, averaged over 500 repeated experiments. The rows represent the dense and sparse signal scenarios and the columns represent different dimensions.}
    \label{fig:exp_errors_block}
\end{figure}

Figure \ref{fig:exp_errors_block} presents the clustering and signal recovery results under the block signal settings in Table \ref{tab:sim_settings}. From Figure \ref{fig:exp_errors_block}, the proposed CFA-PCA and MA-PCA methods much outperformed the three existing methods without considering the block structures. MA-PCA performed notably better in the dense regime, while CFA-PCA excelled in the sparse regime. This trend became more pronounced as the dimension increased, consistent to our theoretical findings.
Spectral clustering performed reasonably well under the dense signal setting and $p_1 = p_2 = 50$, but still lagged behind the two proposed algorithms, and its losses were much larger under the other settings. Also see an illustration of the identified components with nonzero means by the proposed methods in Figure \ref{supp-fig:Example} in the SM.
Those results provide the evidence that the proposed MA-PCA and CFA-PCA methods are superior over the existing methods for clustering and signal recovery under block signals, and the two methods are respectively better suited for dense and sparse block signals.

\section{Case study}\label{sec:data}

\begin{figure}[h]
    \centering
    \includegraphics[width = \textwidth]{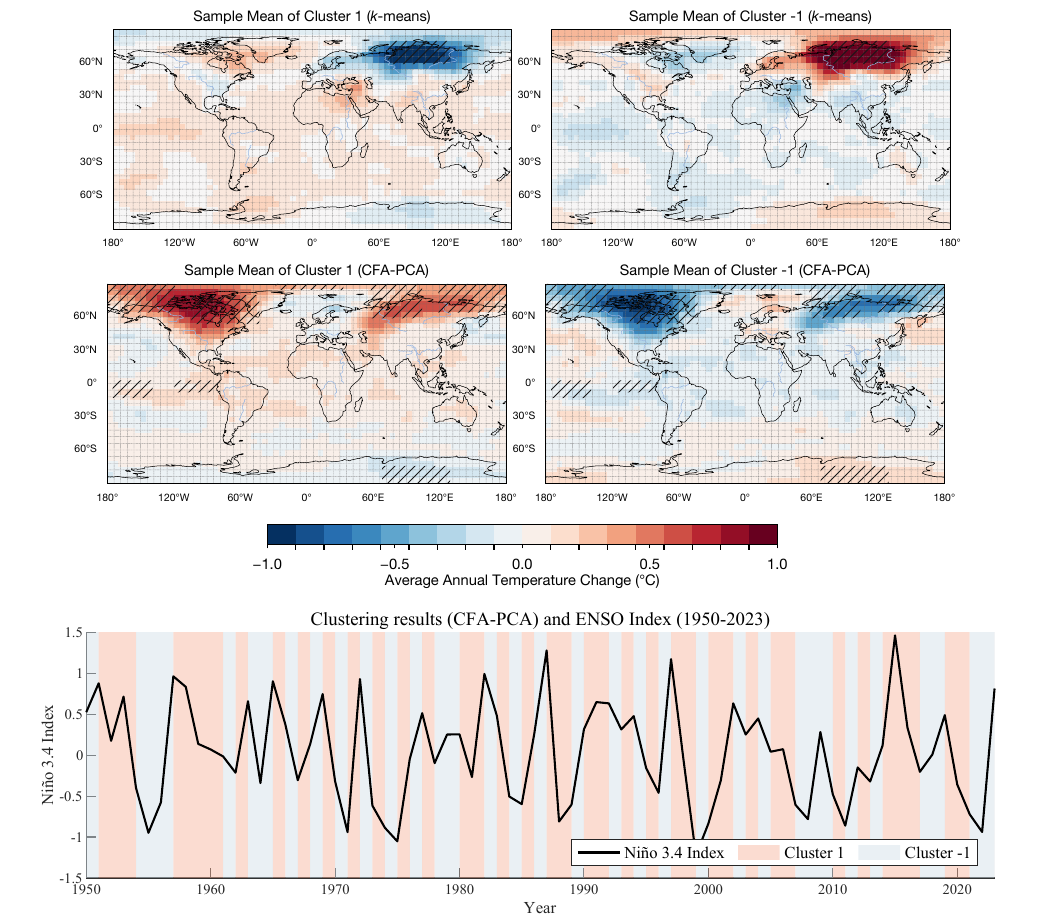}
    \caption{The estimated cluster means of the yearly surface temperature change by $k$-means (upper two panels) and CFA-PCA (middle two panels). The shaded areas indicate the identified signal blocks. 
    The bottom panel shows the clustering results of CFA-PCA and the annual Niño 3.4 index (black line), where the red and gray shading represent the years assigned to the two clusters. 
    }
    \label{fig:Casestudy}
\end{figure}

We study the yearly change of global temperatures to investigate the global warming patterns over time. We extracted the global annual averages of surface temperature anomalies from the HadCRUT5 dataset \citep{HadCRUT5} for the period of 1949--2023, with a spatial resolution of \(5^\circ \times 5^\circ\).
We calculated the yearly differences, resulting in a dataset of 74 observations with the grid size (dimension) of \(72 \times 36 = 2592\).

We applied the CFA-PCA, MA-PCA, IF-PCA, spectral clustering and $k$-means methods to the data for signal recovery and clustering, categorizing the years into two clusters. 
Figure \ref{fig:Casestudy} reports the results of $k$-means (upper two panels) and CFA-PCA (middle two panels). The results of the other three methods were reported in the SM. Of the 74 samples of yearly temperature changes, CFA-PCA assigned 39 to one cluster, and 35 to the other cluster. From the signal recovery map shown in Figure \ref{fig:Casestudy}, CFA-PCA mainly identified four blocks: northern North America, the Siberia region, the equatorial part of the eastern Pacific, and the eastern hemisphere portion of Antarctica. The first three regions exhibit significant warming signals in the first cluster and cooling signals in the second cluster, while the last block shows the opposite pattern. The two clusters identified by the $k$-means algorithm 
appeared to be dominated by the strong signals over northern Siberia. The means of the two clusters exhibited a completely opposite pattern in the Canadian region and northern Siberia. However, these two regions, collectively known as the boreal forests, share very similar climate patterns. 
This shows the results by $k$-means might be unreasonable. 

Specifically, the region 5°N-5°S, 120°-170°W, known as the Niño 3.4 region, contains very weak signals that were successfully detected only by the CFA-PCA algorithm. 
The bottom panel of Figure \ref{fig:Casestudy} shows the annual clustering results of CFA-PCA, where the red and gray shadings represent years assigned to the two clusters. The black line plots the annual Niño 3.4 index. We see that the cluster in red largely matched with the peaks of the Niño 3.4 index, while the other cluster matched with the valleys of this index. This suggests that El Niño and La Niña phenomena might be the intrinsic drivers of the two clusters. These findings further highlight the importance of considering block signals in datasets from Earth sciences and related fields.

\bibliographystyle{apalike}
\bibliography{sample}

\begin{thebibliography}{}

\bibitem[Abbe et~al., 2022]{Abbe2022}
Abbe, E., Fan, J., and Wang, K. (2022).
\newblock An {$\ell_p$} theory of {PCA} and spectral clustering.
\newblock {\em Ann. Statist.}, 50(4):2359--2385.

\bibitem[Arias-Castro et~al., 2005]{Arias-Castro2005}
Arias-Castro, E., Donoho, D.~L., and Huo, X. (2005).
\newblock {Near-optimal detection of geometric objects by fast multiscale
  methods}.
\newblock {\em IEEE Trans. Inform. Theory}, 51(7):2402--2425.

\bibitem[Bickel and Levina, 2008]{Bickel2008}
Bickel, P.~J. and Levina, E. (2008).
\newblock Regularized estimation of large covariance matrices.
\newblock {\em Ann. Statist.}, 36(1):199--227.

\bibitem[Cai et~al., 2023]{Cai2023}
Cai, T.~T., Guo, Z., and Ma, R. (2023).
\newblock Statistical inference for high-dimensional generalized linear models
  with binary outcomes.
\newblock {\em J. Amer. Statist. Assoc.}, 118(542):1319--1332.

\bibitem[Cai et~al., 2019]{Cai2019}
Cai, T.~T., Ma, J., and Zhang, L. (2019).
\newblock C{HIME}: clustering of high-dimensional {G}aussian mixtures with {EM}
  algorithm and its optimality.
\newblock {\em Ann. Statist.}, 47(3):1234--1267.

\bibitem[Cai and Zhang, 2018]{Cai2018}
Cai, T.~T. and Zhang, A. (2018).
\newblock Rate-optimal perturbation bounds for singular subspaces with
  applications to high-dimensional statistics.
\newblock {\em Ann. Statist.}, 46(1):60--89.

\bibitem[Cai et~al., 2010]{Cai2010}
Cai, T.~T., Zhang, C.-H., and Zhou, H.~H. (2010).
\newblock Optimal rates of convergence for covariance matrix estimation.
\newblock {\em Ann. Statist.}, 38(4):2118--2144.

\bibitem[Chen et~al., 2024]{Chen2024}
Chen, P., Chen, A., Yin, S., Li, Y., and Liu, J. (2024).
\newblock {Clustering the Diurnal Cycle of Precipitation Using Global Satellite
  Data}.
\newblock {\em Geophysical Research Letters}, 51(23).

\bibitem[Chen et~al., 2023]{Chen2023}
Chen, S.~X., Qiu, Y., and Zhang, S. (2023).
\newblock Sharp optimality for high-dimensional covariance testing under sparse
  signals.
\newblock {\em Ann. Statist.}, 51(5):1921--1945.

\bibitem[Donoho and Jin, 2004]{Donoho2004}
Donoho, D. and Jin, J. (2004).
\newblock {Higher criticism for detecting sparse heterogeneous mixtures}.
\newblock {\em Ann. Statist.}, 32(3):962--994.

\bibitem[Feng et~al., 2012]{Feng2012}
Feng, D., Tierney, L., and Magnotta, V. (2012).
\newblock {M{RI} tissue classification using high-resolution {B}ayesian hidden
  {M}arkov normal mixture models}.
\newblock {\em J. Amer. Statist. Assoc.}, 107(497):102--119.

\bibitem[Hopkins, 2018]{Hopkins2018}
Hopkins, S. (2018).
\newblock {\em {Statistical Inference and the Sum of Squares Method}}.
\newblock {Ph.D. Thesis}.

\bibitem[Huang et~al., 2023]{Huang2023}
Huang, S., Weng, H., and Feng, Y. (2023).
\newblock Spectral clustering via adaptive layer aggregation for multi-layer
  networks.
\newblock {\em J. Comput. Graph. Statist.}, 32(3):1170--1184.

\bibitem[Ingster, 1997]{Ingster1997}
Ingster, Y.~I. (1997).
\newblock {S}ome problems of hypothesis testing leading to infinitely divisible
  distributions.
\newblock {\em Math. Methods Statist.}, 6(2):266.

\bibitem[Jeng et~al., 2010]{Jeng2010}
Jeng, X.~J., Cai, T.~T., and Li, H. (2010).
\newblock {Optimal sparse segment identification with application in copy
  number variation analysis}.
\newblock {\em J. Amer. Statist. Assoc.}, 105(491):1156--1166.

\bibitem[Jin, 2015]{jin2015}
Jin, J. (2015).
\newblock Fast community detection by score.
\newblock {\em The Annals of Statistics}, 43(1):57--89.

\bibitem[Jin et~al., 2017]{Jin2017}
Jin, J., Ke, Z.~T., and Wang, W. (2017).
\newblock {Phase transitions for high dimensional clustering and related
  problems}.
\newblock {\em Ann. Statist.}, 45(5):2151--2189.

\bibitem[Jin and Wang, 2016]{Jin2016}
Jin, J. and Wang, W. (2016).
\newblock Influential features {PCA} for high dimensional clustering.
\newblock {\em Ann. Statist.}, 44(6):2323--2359.

\bibitem[Kou and Walther, 2022]{Kou2022}
Kou, J. and Walther, G. (2022).
\newblock {Large-scale inference with block structure}.
\newblock {\em Ann. Statist.}, 50(3):1541--1572.

\bibitem[Kunisky et~al., 2022]{Kunisky2022}
Kunisky, D., Wein, A.~S., and Bandeira, A.~S. (2022).
\newblock {Notes on computational hardness of hypothesis testing: predictions
  using the low-degree likelihood ratio}.
\newblock In {\em Mathematical analysis, its applications and computation},
  volume 385 of {\em Springer Proc. Math. Stat.}, pages 1--50. Springer, Cham.

\bibitem[Lei et~al., 2024]{Lei2023}
Lei, J., Zhang, A.~R., and Zhu, Z. (2024).
\newblock {Computational and statistical thresholds in multi-layer stochastic
  block models}.
\newblock {\em Ann. Statist.}, 52(5):2431--2455.

\bibitem[Liu et~al., 2008]{Liu2008}
Liu, Y., Hayes, D.~N., Nobel, A., and Marron, J.~S. (2008).
\newblock Statistical significance of clustering for high-dimension,
  low–sample size data.
\newblock {\em Journal of the American Statistical Association},
  103(483):1281--1293.

\bibitem[L\"offler et~al., 2022]{Loffler2022}
L\"offler, M., Wein, A.~S., and Bandeira, A.~S. (2022).
\newblock Computationally efficient sparse clustering.
\newblock {\em Inf. Inference}, 11(4):1255--1286.

\bibitem[L\"offler et~al., 2021]{Loffler2021}
L\"offler, M., Zhang, A.~Y., and Zhou, H.~H. (2021).
\newblock Optimality of spectral clustering in the {G}aussian mixture model.
\newblock {\em Ann. Statist.}, 49(5):2506--2530.

\bibitem[Lyu and Xia, 2023]{Lyu2023}
Lyu, Z. and Xia, D. (2023).
\newblock Optimal estimation and computational limit of low-rank {G}aussian
  mixtures.
\newblock {\em Ann. Statist.}, 51(2):646--667.

\bibitem[Morice et~al., 2021]{HadCRUT5}
Morice, C.~P., Kennedy, J.~J., Rayner, N.~A., Winn, J.~P., Hogan, E., Killick,
  R.~E., Dunn, R. J.~H., Osborn, T.~J., Jones, P.~D., and Simpson, I.~R.
  (2021).
\newblock {An Updated Assessment of Near-Surface Temperature Change From 1850:
  The HadCRUT5 Data Set}.
\newblock {\em Journal of Geophysical Research: Atmospheres}, 126(3).

\bibitem[Qiu and Guo, 2024]{qiu-2024}
Qiu, Y. and Guo, B. (2024).
\newblock Minimax detection boundary and sharp optimal test for gaussian
  graphical models.
\newblock {\em Journal of the Royal Statistical Society Series B: Statistical
  Methodology}, 86(5):1221--1242.

\bibitem[Schramm and Wein, 2022]{Schramm2022}
Schramm, T. and Wein, A.~S. (2022).
\newblock Computational barriers to estimation from low-degree polynomials.
\newblock {\em Ann. Statist.}, 50(3):1833--1858.

\bibitem[Sun et~al., 2024]{Sun2024}
Sun, H.-X., Wang, S., Zheng, X., and Chen, S.~X. (2024).
\newblock High-dimensional ensemble kalman filter with localization, inflation,
  and iterative updates.
\newblock {\em Quarterly Journal of the Royal Meteorological Society},
  150(765):4870--4884.

\bibitem[Sun et~al., 2012]{Sun2012}
Sun, W., Wang, J., and Fang, Y. (2012).
\newblock Regularized k-means clustering of high-dimensional data and its
  asymptotic consistency.
\newblock {\em Electron. J. Stat.}, 6:148--167.

\bibitem[Zhang and Zhou, 2024]{Zhang2024}
Zhang, A.~Y. and Zhou, H.~Y. (2024).
\newblock Leave-one-out singular subspace perturbation analysis for spectral
  clustering.
\newblock {\em Ann. Statist.}, 52(5):2004--2033.

\bibitem[Zou and Xue, 2018]{Zou2018}
Zou, H. and Xue, L. (2018).
\newblock A selective overview of sparse principal component analysis.
\newblock {\em Proceedings of the IEEE}, 106(8):1311--1320.

\end{thebibliography}


\begin{thebibliography}{}

\bibitem[Abbe et~al., 2022]{Abbe2022}
Abbe, E., Fan, J., and Wang, K. (2022).
\newblock An {$\ell_p$} theory of {PCA} and spectral clustering.
\newblock {\em Ann. Statist.}, 50(4):2359--2385.

\bibitem[Ahle, 2022]{Ahle2022}
Ahle, T.~D. (2022).
\newblock Sharp and simple bounds for the raw moments of the binomial and
  {P}oisson distributions.
\newblock {\em Statist. Probab. Lett.}, 182:Paper No. 109306, 5.

\bibitem[Aldous, 1985]{Aldous1985}
Aldous, D.~J. (1985).
\newblock Exchangeability and related topics.
\newblock In {\em \'Ecole d'\'et\'e{} de probabilit\'es de {S}aint-{F}lour,
  {XIII}---1983}, volume 1117 of {\em Lecture Notes in Math.}, pages 1--198.
  Springer, Berlin.

\bibitem[Arias-Castro et~al., 2005]{Arias-Castro2005}
Arias-Castro, E., Donoho, D.~L., and Huo, X. (2005).
\newblock {Near-optimal detection of geometric objects by fast multiscale
  methods}.
\newblock {\em IEEE Trans. Inform. Theory}, 51(7):2402--2425.

\bibitem[Cai et~al., 2011]{Cai2011}
Cai, T.~T., Jeng, X.~J., and Jin, J. (2011).
\newblock Optimal detection of heterogeneous and heteroscedastic mixtures.
\newblock {\em J. R. Stat. Soc. Ser. B Stat. Methodol.}, 73(5):629--662.

\bibitem[Cai and Zhang, 2018]{Cai2018}
Cai, T.~T. and Zhang, A. (2018).
\newblock Rate-optimal perturbation bounds for singular subspaces with
  applications to high-dimensional statistics.
\newblock {\em Ann. Statist.}, 46(1):60--89.

\bibitem[Chen et~al., 2021]{Spect}
Chen, Y., Chi, Y., Fan, J., and Ma, C. (2021).
\newblock {Spectral Methods for Data Science: A Statistical Perspective}.
\newblock {\em Foundations and Trends® in Machine Learning}, 14(5):566–806.

\bibitem[Chen et~al., 2024]{ChenY2024}
Chen, Y., Zhao, C., and Zhi, H. (2024).
\newblock {Analysis of ENSO Event Intensity Changes and Time–Frequency
  Characteristic Since 1875}.
\newblock {\em Atmosphere}, 15(12).

\bibitem[Donoho and Jin, 2004]{Donoho2004}
Donoho, D. and Jin, J. (2004).
\newblock {Higher criticism for detecting sparse heterogeneous mixtures}.
\newblock {\em Ann. Statist.}, 32(3):962--994.

\bibitem[Hall and Jin, 2010]{Hall2010}
Hall, P. and Jin, J. (2010).
\newblock Innovated higher criticism for detecting sparse signals in correlated
  noise.
\newblock {\em Ann. Statist.}, 38(3):1686--1732.

\bibitem[Jeng et~al., 2010]{Jeng2010}
Jeng, X.~J., Cai, T.~T., and Li, H. (2010).
\newblock {Optimal sparse segment identification with application in copy
  number variation analysis}.
\newblock {\em J. Amer. Statist. Assoc.}, 105(491):1156--1166.

\bibitem[Jin et~al., 2017]{Jin2017}
Jin, J., Ke, Z.~T., and Wang, W. (2017).
\newblock {Phase transitions for high dimensional clustering and related
  problems}.
\newblock {\em Ann. Statist.}, 45(5):2151--2189.

\bibitem[Kou and Walther, 2022]{Kou2022}
Kou, J. and Walther, G. (2022).
\newblock {Large-scale inference with block structure}.
\newblock {\em Ann. Statist.}, 50(3):1541--1572.

\bibitem[Kunisky et~al., 2022]{Kunisky2022}
Kunisky, D., Wein, A.~S., and Bandeira, A.~S. (2022).
\newblock {Notes on computational hardness of hypothesis testing: predictions
  using the low-degree likelihood ratio}.
\newblock In {\em Mathematical analysis, its applications and computation},
  volume 385 of {\em Springer Proc. Math. Stat.}, pages 1--50. Springer, Cham.

\bibitem[Lei et~al., 2024]{Lei2023}
Lei, J., Zhang, A.~R., and Zhu, Z. (2024).
\newblock {Computational and statistical thresholds in multi-layer stochastic
  block models}.
\newblock {\em Ann. Statist.}, 52(5):2431--2455.

\bibitem[L\"offler et~al., 2022]{Loffler2022}
L\"offler, M., Wein, A.~S., and Bandeira, A.~S. (2022).
\newblock Computationally efficient sparse clustering.
\newblock {\em Inf. Inference}, 11(4):1255--1286.

\bibitem[Petrov, 1995]{Petrov1995}
Petrov, V.~V. (1995).
\newblock {\em Limit theorems of probability theory}, volume~4 of {\em Oxford
  Studies in Probability}.
\newblock The Clarendon Press, Oxford University Press, New York.
\newblock Sequences of independent random variables, Oxford Science
  Publications.

\bibitem[Rayner et~al., 2003]{NINO34}
Rayner, N.~A., Parker, D.~E., Horton, E.~B., Folland, C.~K., Alexander, L.~V.,
  Rowell, D.~P., Kent, E.~C., and Kaplan, A. (2003).
\newblock Global analyses of sea surface temperature, sea ice, and night marine
  air temperature since the late nineteenth century.
\newblock {\em Journal of Geophysical Research: Atmospheres}, 108(D14).

\bibitem[Rudelson and Vershynin, 2013]{Rudelson2013}
Rudelson, M. and Vershynin, R. (2013).
\newblock Hanson-{W}right inequality and sub-{G}aussian concentration.
\newblock {\em Electron. Commun. Probab.}, 18:no. 82.

\bibitem[van~der Vaart, 1998]{Varrt1998}
van~der Vaart, A.~W. (1998).
\newblock {\em {Asymptotic statistics}}, volume~3 of {\em Cambridge Series in
  Statistical and Probabilistic Mathematics}.
\newblock Cambridge University Press, Cambridge.

\bibitem[Vershynin, 2011]{vershynin2011}
Vershynin, R. (2011).
\newblock Introduction to the non-asymptotic analysis of random matrices.

\bibitem[Vershynin, 2018]{HDP-BOOK}
Vershynin, R. (2018).
\newblock {\em High-dimensional probability}, volume~47 of {\em Cambridge
  Series in Statistical and Probabilistic Mathematics}.
\newblock Cambridge University Press, Cambridge.

\bibitem[Zhang and Chen, 2021]{concentration}
Zhang, H. and Chen, S. (2021).
\newblock Concentration inequalities for statistical inference.
\newblock {\em Communications in Mathematical Research}, 37(1):1--85.

\end{thebibliography}

\end{document}